\documentstyle[aps,eqsecnum]{revtex}
\begin{document}
\baselineskip=16pt
\title{Scenario for Ultrarelativistic Nuclear Collisions: \\
  III.~ Gluons in the expanding geometry. }            
\author{ A. Makhlin}
\address{Department of Physics and Astronomy, Wayne State University, 
Detroit, MI 48202}
\date{July 26, 2000}
\maketitle
\begin{abstract}
We derive  expressions for  various correlators of the gauge  field and find
the propagators in Hamiltonian dynamics which employs a new gauge $A^\tau=0$.
This gauge is a part of the wedge form of relativistic dynamics suggested
earlier  as a tool for the study of quantum dynamics in ultra-relativistic
heavy ion collisions. We prove that the gauge is completely fixed. The  gauge
field is quantized and the field of radiation and the longitudinal fields are
unambiguously separated. The new gauge  puts the quark and gluon fields of the
colliding hadrons in one Hilbert space and thus allows one to avoid 
factorization.                                                         
\end{abstract}

\section{Introduction}\label{sec:SN1}

In two  previous papers of this cycle \cite{tev,gqm}( further quoted as papers
[I] and [II]), we   explained the physical motivation of the ``wedge form of 
dynamics'' as a promising tool to explore the processes which take place
during
the collision of two heavy ions. In compliance with the general definition of
dynamics given by  Dirac \cite{Dirac}, the wedge form  includes a specific
definition of  the quantum mechanical observables on space-like surfaces, as
well as the means to describe the evolution of the observables  from an
``earlier'' space-like surface to a ``later'' one.  Unlike the other forms,
the
wedge form explicitly refers to the two main geometrical features of the
phenomenon, {\em i.e.,} the strong localization of the initial interaction
and,
as a consequence, the absence of  translational invariance in the temporal and
longitudinal directions.

In the wedge form of dynamics, the states of the quark and gluon fields are
defined on  the space-like hyper-surfaces of the constant proper  time $\tau$,
$\tau^2=t^2-z^2$.  The states of  fermion fields were discussed
in paper [II]. In this paper, we continue the study of the gluons and augment
our previous consideration by the gauge condition $A^\tau =0$. This simple
idea
solves several problems.  First, this gauge  condition is boost-invariant and
thus complies with the symmetry of the collision. Second, it becomes  possible
to treat two different light-front gauges (which describe  gluons from each
nucleus of the initial state separately) as  the two limits of this single
gauge. Therefore, the new approach keeps important  connections with the
theory
of  deep inelastic $ep$-scattering (DIS). This fact is vital for the
subsequent
calculations since $e-p$ DIS is the only existing source
of data on  nucleon structure in  high-energy collisions. The approach
based on quantum field kinetics (QFK) allows one to treat both the nuclear
collision and $e-p$ DIS as the similar transient processes. Third, after the
collision, this kind of gauge becomes a  local temporal axial gauge, thus
providing a smooth transition to the Bjorken regime of the boost-invariant
expansion.  

Most of this paper is technical, and any relevant physical discussion  of the
results appears only after their mathematical derivation. These results
were summarized in paper [II]. Since the first interaction of two {\em
finite-sized} nuclei is strongly localized, the geometrical symmetry of the
final state is manifestly broken and the observables of wedge dynamics are
essentially defined on the curved space-like hypersurface. For the fermion
field this have led to an obvious Thomas precession. Similar orientation
effects happen in the case
of the vector field also. The dynamics of the gauge field is rich, and the
procedure of its quantization triggers many puzzles that can be traced back
to the classical roots of the gauge field theory.

In section \ref{subsec:SB21},  we derive equations of motion for the gauge
field in the gauge $A^\tau =0$, find the Hamiltonian variables and the
normalization
condition. The equations of motion are linearized and the modes of the free 
radiation field are obtained in section \ref{subsec:SB22}. In section
\ref{subsec:SB23}, the retarded propagator of the perturbation theory is found
as the response function of the field on the external current. This part of
the
calculation turned out to be the most durable, since the gauge
condition is inhomogeneous and none of the modern  methods is
effective. However, the old-fashioned variation of parameters does  work.  
The most important result of this paper, separation the transverse and
longitudinal parts of the gluon propagator is obtained here. We essentially
base calculation of the quark self-energy in expanding quark-gluon system on
this result. These calculations are presented in the next paper. In section
\ref{subsec:SB25}, we show that the previously obtained propagator solves the
initial data problem for the gauge field. Unlike in the homogeneous axial
gauges, the propagators of the gauge $A^\tau=0$ do not have any spurious
poles. 

Section \ref{sec:SN3} is devoted to the quantization of the vector field in
the gauge $A^\tau=0$. We begin in \ref{subsec:SB30} with the proof  of the
fact that the gauge $A^\tau=0$ can be completely fixed provided the
physical charge density, $\tau j_\tau$ vanishes at $\tau=0$, the moment of
the first touch of the nuclei. This is exactly what can be expected from
the nuclei colorlessness. Then, the Gauss law can be unquestionably
used to eliminate the unphysical degrees of freedom in the equations of
motion. We continue in \ref{subsec:SB31} with a computation of the Wightman
functions, and  study  the causal  properties of the commutators in
\ref{subsec:SB32}. The latter appears to be abnormal; the Riemann function is
not symmetric and penetrates the exterior of the light cone. However, the
behavior of the observables is fully causal and the  procedure of the
canonical quantization is accomplished in \ref{subsec:SB33}.  Even though it
is impossible to introduce transverse and longitudinal currents (as it is
customary for the homogeneous gauge conditions) and thus fully
separate the dynamics of the corresponding fields, we found it useful to
discriminate the various field patterns by the type of their propagation. The
propagator of the  transverse field is sensitive to the light cone boundaries
while the longitudinal and instantaneous parts of the field do not propagate.
These coordinate form of these two fragments of the response function is
derived in section \ref{sec:SN5}. Ultimately, the longitudinal part of the
gluon propagator appeared to be of the greatest importance for the dynamics
of the screening effects at the early stage of the collision.

In Appendix~4, we study the limiting behavior of the propagator
in the central rapidity region and in the vicinity of the null-planes and show
that propagators of the gauges $A^0=0$ and  $A^\pm=0$ are recovered.   It is
important that the spurious poles are recovered only in the unphysical limit
of infinite rapidity. This result is of practical importance because it
establishes the connection between the new approach and the existing theory of
the deeply  inelastic processes at high energies.

\section{The classical treatment}
\label{sec:SN2}                

The final goal of this paper is to build a quantum theory of the vector gauge
field in the expanding geometry of nuclear collision. Development of a quantum
theory always begins with its classical counterpart which provides the
one-particle wave functions (which later serve as quantum states) and the
classical Green functions (which later become the propagators of quantum
theory).  Furthermore, the quantum propagator of gauge field includes
the longitudinal part which can be found only by classical analysis.
The classical part of this program is the subject of this section.

\subsection{Classical equations of motion}
\label{subsec:SB21}  

We now consider the case of pure glue-dynamics. We denote  the gluon field in
the fundamental representation of the color group as $A_\mu(x)=t^a
A^{a}_{\mu}(x)$.  Consequently, we have the field tensor,
\begin{eqnarray}  
F_{\mu\nu}=t^a F^{a}_{\mu\nu}={\cal D}_\mu A_\nu-{\cal D}_\nu A_\mu=
\partial_\mu A_\nu -\partial_\nu A_\mu - ig [A_\mu,A_\nu], \nonumber
\end{eqnarray}
where ${\cal D}_\mu= \partial_\mu -ig[A_\mu(x), ...]$ is the covariant
derivative on the local color group.
The gauge invariant action of the theory looks as follows,
\begin{eqnarray}
{\cal S}=\int {\cal L}(x)d^4x =\int [-{1\over 4}
{\rm g}^{\mu\lambda}(x){\rm g}^{\nu\sigma}(x)F_{\mu\nu}(x)F_{\lambda\sigma}(x)
-j^\mu A_\mu ]\sqrt{-{\rm g}} d^4 x.
\label{eq:E2.1}\end{eqnarray}       
Its variation with respect to the gluon field yields the Lagrangian
equations of motion,
\begin{eqnarray}
\partial_\lambda [(-{\rm g})^{1/2}{\rm g}^{\mu\lambda}
{\rm g}^{\nu\sigma}F_{\mu\nu}] - 
ig (-{\rm g})^{1/2} [A_\lambda, {\rm g}^{\mu\lambda}
{\rm g}^{\nu\sigma}F_{\mu\nu}]=
(-{\rm g})^{1/2}j^\sigma~~,
\label{eq:E2.2}\end{eqnarray}
where $j^\mu$ is the color current of the fermion fields and $~~{\rm g}=
{\rm det}|{\rm g}_{\mu\nu}|$. The equations are twice covariant, 
{\em i.e.}, with
respect to the gauge transformations in color space and the arbitrary
transformations of the coordinates. In what follows, we shall employ the
special coordinates associated with the constant proper time  hyper-surfaces
inside the light cone of the collision point $t=z=0$. The new coordinates
parameterize the Minkowski coordinates $(t,x,y,z)$ as
$(\tau\cosh\eta,x,y,\tau\sinh\eta)$. In addition, we impose the gauge 
condition $A_\tau =0$.     The corresponding gauge transformation is well
defined.  Indeed, let $A_{\mu}(x)$ be an arbitrary field configuration and
$A'_{\mu}(x)$ its gauge transform with the generator 
\begin{eqnarray}
U(\tau,\eta, \vec{r_\bot})=
P_\tau \exp \{-\int_{0}^{\tau} A_\tau (\tau',\eta, \vec{r_\bot})d\tau'\},
\label{eq:E2.6}\end{eqnarray}   
then the new field, $A'_{\mu}=U A_{\mu}U^{-1} + \partial_\mu UU^{-1}$, obeys
the condition $A^\tau=0$.    
Imposing this gauge condition we arrive at the system of four equations:
\begin{eqnarray}
{\cal C}(x)={1\over \tau}\partial_\eta \partial_\tau A_\eta + \tau
\partial_r \partial_\tau A_r -ig\{{1\over\tau}[A_\eta,\partial_\tau A_\eta]
+\tau [A_r, \partial_\tau A_r]\} -\tau j^\tau = 0,
\label{eq:E2.3}\end{eqnarray}       
\begin{eqnarray}
-\partial_\tau \tau \partial_\tau A_r +
{1\over \tau}\partial_\eta(\partial_\eta A_r-\partial_r A_\eta)
 + \tau \partial_s (\partial_s A_r-\partial_r A_s) \nonumber \\
- ig\{{1\over \tau}\partial_\eta [A_\eta ,A_r] +\tau\partial_s[A_s,A_r]+
{1\over \tau}[A_\eta,F_{\eta r}] + \tau [A_s,F_{sr}]\} 
-\tau j^r =0~,
\label{eq:E2.4}\end{eqnarray}
\begin{eqnarray}
-\partial_\tau {1\over \tau} \partial_\tau A_\eta+ 
{1\over \tau}\partial_r (\partial_r A_\eta-\partial_\eta A_r)-
ig \big[{1\over \tau}\partial_r[A_r,A_\eta]
+{1\over\tau}[A_r,F_{r\eta}] \big]-\tau j^\eta =0
\label{eq:E2.5}\end{eqnarray}  
Here, we  use the Latin indices from $r$ to $w$  for the transverse $x$- and
$y$-components ($r,...,w=1,2$). We shall also use the arrows over the
letters to denote the  two-dimensional vectors, {\em e.g.}, 
${\vec k}=(k_x,k_y)$,
$|{\vec k}|=k_{\bot}$. The Latin indices from $i$ to $n$ ($i,...,n=1,2,3$)
will be used  for the three-dimensional internal coordinates $u^i=
(x,y,\eta)$ on the hyper-surface $\tau=const$. The metric tensor has only
diagonal components ${\rm g}_{\tau\tau}=-{\rm g}_{xx}= -{\rm g}_{yy}=1$,
${\rm g}_{\eta\eta}= -\tau^2$. The first of these equations (\ref{eq:E2.3})
contains no
second order time derivatives  and is a constraint rather than a dynamical
equation.  The constraint weakly equals to zero in classical Hamiltonian
dynamics and serves as a condition 
imposed on physical states in the quantum theory. 
The canonical momenta of the theory are as follows: 
\begin{eqnarray}
p^\tau=0,\;\;\;\; p^\eta={1\over \tau} F_{\tau\eta}= {1\over \tau}
\stackrel{\bullet}{A}_\eta,\;\;\;\;  
p_r=\tau F_{\tau r}=\tau\stackrel{\bullet}{A}_r~.
\label{eq:E2.7}\end{eqnarray} 
Hereafter, the dot above the letter denotes a derivative with respect to the
Hamiltonian time $\tau$.  Because of the gauge condition, the canonical
momenta do not contain the color commutators. After excluding the velocities,
the Hamiltonian can be written down  in the canonical variables, 
\begin{eqnarray} 
H=\int d\eta d \vec{r_\bot}
\tau\{{1\over 2} p^\eta p^\eta + {1\over 2\tau^2} p^r p^r +{1\over
2\tau^2} F_{\eta r}F_{\eta r}+ {1\over 4}F_{rs}F_{rs} +j^\eta A_\eta+ j^r
A_r\} ~.
\label{eq:E2.8}\end{eqnarray} 
Then the equations (\ref{eq:E2.4}) and (\ref{eq:E2.5}) are immediately
recognized as the Hamiltonian equations of motion. The Poisson bracket of
the constraint ${\cal C}$ with the Hamiltonian vanishes, thus creating the
generator of the residual gauge transformations which are tangent to the
hyper-surface.  Conservation of the constraint is a direct consequence of
the Lagrange (or Hamiltonian) classical  equations of motion as well.

The normalization condition for the one-particle solutions is obviously 
derived from the charge conservation law. For the gauge field, this is
impossible. Therefore, we shall accept the condition which supports
self-adjointness of the homogeneous system after its linearization.
This leads to a natural definition for the scalar product of the states 
of the  vector field in the gauge $A^\tau =0$~,       
\begin{eqnarray}
(V,W)=\int_{-\infty}^{\infty}d\eta\int d^2{\vec r} \tau {\sf g}^{ik}
V^{*}_{i} i{\stackrel{\leftrightarrow}{\partial}}_{\tau} W_k ~,
\label{eq:E2.14}\end{eqnarray}               
where ${\sf g}^{ik}$ is the metric tensor of the three-dimensional 
internal geometry of the hyper-surface $\tau = const$~.  This norm of
the one-particle states prevents them from flowing out of the interior
of the past and future light wedges of the interaction plane.

\subsection{Modes of the free radiation field. Field of the static source}
\label{subsec:SB22}  

As a tool for the future development of the perturbation theory, we need to
find the propagators and Wightman functions when the nonlinear self-interaction
of the gluon field is switched off. In this case, the system of  equations for
the non-vanishing components of the vector potential  and the constraint look
as follows,
\begin{eqnarray}
[\partial_{\tau}\tau\partial_{\tau}-{1\over \tau }\partial_{\eta}^{2} 
-\tau\partial_{s}^{2}]A_r +\partial_{r}[\tau\partial_{s}A_s + 
{1 \over \tau}\partial_{\eta}A_\eta] =-\tau j^r~~,
\label{eq:E2.10}\end{eqnarray}         
\begin{eqnarray}
\big[\partial_{\tau}{1\over\tau} \partial_{\tau} -
{1\over\tau} \partial_{s}^{2}\big] A_\eta
+{1\over \tau}\partial_{\eta}\partial_{s}A_s =-\tau j^\eta~~,
\label{eq:E2.11}\end{eqnarray}
\begin{eqnarray}
{\cal C}(x)={1\over \tau}\partial_\eta \partial_{\tau}A_\eta+
\tau\partial_{\tau} \partial_{r}A_r -\tau j^\tau= 0~,
\label{eq:E2.13}\end{eqnarray}
where $j^\mu$ includes all kinds of the color currents. An explicit form of the
solution for the homogeneous system is found in Appendix 1. In compliance with
the gauge condition (which explicitly eliminates  one of four field components)
we  find three modes $V^{(\lambda)}$  of the free vector field. Two transverse
modes obey Gauss' law without the charge and have the unit norm (see Appendix 1)
with respect to the scalar product (\ref{eq:E2.14}):
\begin{eqnarray}  
V^{(1)}_{{\vec k},\nu}(x)={e^{-\pi\nu/2}\over 2^{5/2}\pi k_{\bot}} 
\left( \begin{array}{c} 
                         k_y \\ 
                        -k_x \\ 
                         0 
                             \end{array} \right)
H^{(2)}_{-i\nu} (k_{\bot}\tau) e^{i\nu\eta +i{\vec k}{\vec r}};\;\;\; 
{\rm and}\;\;\;
V^{(2)}_{{\vec k},\nu}(x)={e^{-\pi\nu/2}\over 2^{5/2}\pi k_{\bot}} 
\left( \begin{array}{c} 
                 \nu k_x R^{(2)}_{-1,-i\nu}(k_{\bot}\tau) \\ 
                 \nu k_y R^{(2)}_{-1,-i\nu}(k_{\bot}\tau) \\ 
                 - R^{(2)}_{1,-i\nu}(k_{\bot}\tau) 
                                            \end{array} \right)
                e^{i\nu\eta +i{\vec k}{\vec r}}~.
\label{eq:E2.15}\end{eqnarray}                                  
The mode $V^{(2)}$ is constructed from the functions 
$R^{(j)}_{\mu,-i\nu}(k_{\bot}\tau)=R^{(j)}_{\mu,-i\nu}(k_{\bot}\tau|s)$
corresponding to the boundary condition of vanishing gauge field at $\tau=0$.
This guarantees continuous behavior of the field at $\tau=0$. Indeed, as
$\tau\rightarrow 0$,~ the normal and the tangential directions become
degenerate. As long as $A^\tau=0$ is the gauge condition,  continuity requires
that  $A^\eta\rightarrow 0$ as $\tau\rightarrow 0$.

It is instructive to know the physical components of the electric and magnetic 
fields of these modes,  $ {\cal E}^m=\sqrt{-{\rm g}}{\rm g}^{mn}
\stackrel{\bullet}{A}_n$  and  $ {\cal B}^m=-(2\sqrt{-{\rm g}})^{-1}
e^{mln}F_{ln}$: 
\begin{eqnarray}  
{\cal E}^{(1)m}_{{\vec k},\nu}(x)=i{\cal B}^{(2)m}_{{\vec k},\nu}(x)
={e^{-\pi\nu/2}\over 2^{5/2}\pi k_{\bot}} 
\left( \begin{array}{c}  k_y \\ 
                        -k_x \\ 
                         0   \end{array} \right)
\stackrel{\bullet}{H}^{(2)}_{-i\nu} (k_{\bot}\tau) 
e^{i\nu\eta +i{\vec k}{\vec r}}~~, 
\nonumber\\
{\cal E}^{(2)m}_{{\vec k},\nu}(x)=i{\cal B}^{(1)m}_{{\vec k},\nu}(x)
={e^{-\pi\nu/2}\over 2^{5/2}\pi k_{\bot}} 
\left( \begin{array}{c}  \nu k_x  \\ 
                         \nu k_y  \\ 
                         - k_{\bot}^{2} \end{array} \right)
H^{(2)}_{-i\nu} (k_{\bot}\tau) e^{i\nu\eta +i{\vec k}{\vec r}}~.
\label{eq:E2.15A}\end{eqnarray}    
The mode $V^{(2)}$ can be obtained from the mode $V^{(1)}$ by a simple
interchange of its electric and magnetic fields. Using  standard wave-guide
terminology, one may call mode $V^{(1)}$  as the ``transverse electric mode''
and the mode $V^{(2)}$  as the ``transverse magnetic mode''.  Equations
(\ref{eq:E2.15A}) indicate, that the field strength tensor of the free
radiation field obeys the condition, $(F^\ast)^\ast =-F$. Therefore, certain
linear combinations of the modes $V^{(1)}$ and $V^{(2)}$ may be analytically
continued to Euclidean space as self-dual solutions of the field equations.

An equivalent full set of the transverse modes carries  (instead of the  boost
$\nu$) the quantum number $\theta$, (rapidity), {\em i.e.} $k_0= k_{\bot}
\cosh\theta,~~k_3= k_{\bot} \sinh\theta$. These functions can be obtained by
means of the Fourier transform,

\begin{eqnarray}  
v^{(\lambda)}_{{\vec k},\theta}(x)
= \int_{-\infty}^{+\infty}  {d\nu \over (2\pi)^{1/2}i} e^{-i\nu\theta}
V^{(\lambda)}_{{\vec k},\nu}(x)~~,
\label{eq:E2.15a} \end{eqnarray}        
and have the following form,
\begin{eqnarray}  
v^{(1)}_{{\vec k},\theta}(x)={1\over 4\pi^{3/2} k_{\bot}} 
\left( \begin{array}{c} 
                         k_y \\ 
                        -k_x \\ 
                         0 
                             \end{array} \right)
e^{-ik_{\bot}\tau\cosh (\theta-\eta) +i{\vec k}{\vec r}}; \;\;\;
v^{(2)}_{{\vec k},\theta}(x)={1\over 4\pi^{3/2} k_{\bot}}  
\left( \begin{array}{c} 
                 k_x f_1 \\ 
                 k_y f_1 \\ 
                  - f_2 
                 \end{array} \right) e^{i{\vec k}{\vec r}}~~,
\label{eq:E2.15b}\end{eqnarray}  
where
\begin{eqnarray}  
f_1(\tau ,\eta)=  k_{\bot} \sinh (\theta-\eta) \int_{0}^{\tau} 
e^{-ik_{\bot}\tau' \cosh (\theta-\eta)}d\tau' =i \tanh (\theta-\eta)
(e^{-ik_{\bot}\tau \cosh (\theta-\eta)} - 1)~~, \nonumber
\end{eqnarray}                                             
\begin{eqnarray}  
f_2(\tau ,\eta)= k_{\bot}^2\int_{0}^{\tau}
e^{-ik_{\bot}\tau' \cosh (\theta-\eta)}\tau' d\tau'=
{e^{-ik_{\bot}\tau \cosh (\theta-\eta)} -1 \over
\cosh^{2} (\theta-\eta)} +
i k_{\bot}\tau {e^{-ik_{\bot}\tau\cosh (\theta-\eta)} 
\over  \cosh (\theta-\eta)}~~.
\label{eq:E2.15c} 
\end{eqnarray}        
The norm of the Coulomb mode $V^{(3)}$, as defined by Eq.~(\ref{eq:E2.14}),
equals zero,
and it is orthogonal to $V^{(1)}$ and $V^{(2)}$. Though this solution 
obeys the equations of motion without the current, it does not obey Gauss'
law without a charge. Therefore, it should be discarded in the decomposition
of the radiation field. However, it should be kept if we consider
the radiation field in the presence of a static source with the
$\tau$-independent density  
$\rho({\vec k},\nu)=\tau j^{\tau}_{{\vec k}\nu}(\tau)=const(\tau) $. In this
case, its definition can be completed using  Gauss' law:
\begin{eqnarray} 
V^{(3)}_{{\vec k},\nu}(x)={ \rho({\vec k},\nu) \over (2\pi)^3 i k_{\bot}^2}
 \left( \begin{array}{c} 
                 k_r  Q_{-1,i\nu}(k_{\bot}\tau) \\ 
                 \nu~Q_{1,i\nu}(k_{\bot}\tau)
                                            \end{array} \right) 
                   e^{i\nu\eta +i{\vec k}{\vec r}}~~.     
\label{eq:E2.16}\end{eqnarray}                     
The coordinate form of this solution is noteworthy. The physical
components, ~${\cal E}^m=\sqrt{-{\rm g}}{\rm g}^{ml}{\stackrel{\bullet} A}_l$~, of
the electric field of the ``$\tau$--static'' source can be written in
the integral form,
\begin{eqnarray}   
{\cal E}_{(stat)}^{i}(\tau,{\vec r}_1,\eta_1)= \int d{\vec r}_2 d\eta_2 
K_i(\tau;{\vec r}_1-{\vec r}_2,\eta_1 -\eta_2) \rho({\vec r}_2,\eta_2)~,
\label{eq:E2.17}\end{eqnarray}   
with the kernel
\begin{eqnarray}   
K_i(\tau;{\vec r},\eta)= \int {d\nu d^2{\vec k}\over (2\pi)^3}
{e^{i\nu\eta +i{\vec k}{\vec r}}\over i k_{\bot}^2} 
\left( \begin{array}{c} 
                 k_r   s_{1,i\nu}(k_{\bot}\tau) \\ 
                 \nu k_{\bot}^2  s_{-1,i\nu}(k_{\bot}\tau)
                                            \end{array} \right) 
              = -{\theta(\tau -r_{\bot})\over 4\pi}
\left( \begin{array}{c} 
                   \tau\cosh\eta(\partial / \partial x^r) \\ 
                    \partial / \partial(\tau\sinh\eta)
                           \end{array} \right) {1\over R_{12}}~~,
\label{eq:E2.18}\end{eqnarray}

where  $~R_{12}=(r_{\bot}^2+\tau^2\sinh^2\eta)^{1/2}= [({\vec r}_1-{\vec
r}_2)^2+\tau^2\sinh^2(\eta_1 -\eta_2)]^{1/2}~$, is the distance between the
points $({\vec r}_1,\eta_1)$ and  $({\vec r}_2,\eta_2)$  in the internal
geometry of the surface $\tau=const$. The technical details of the derivation
of the last expression  will be presented in Sec.~\ref{sec:SN5}.
Eq.~(\ref{eq:E2.18}) is an analog of Coulomb's law of electrostatics, except
that now the source has a  density which is static with respect to the
Hamiltonian time $\tau$. In fact, {\em the source is static if  it expands in
such a way that its physical component}  $~{\cal J}^\tau=\tau j^\tau(\tau,
\eta, {\vec r})~$ {\em does not depend on} $~\tau$.  These expressions  will be
helpful in  recognizing the origin of  various terms in the full propagator
which is calculated below.

\subsection{Propagator in the gauge $A^\tau=0$}
\label{subsec:SB23}  

The calculation of the propagator in the gauge $A^\tau=0$ (associated with the
system of the curved surfaces $\tau=const$)  meets several problems. Three
methods are commonly used in  field theory. One of them strongly appeals to the
Fourier analysis in the plane Minkowski space which is not applicable now
because the metric itself is coordinate-dependent. The second method uses the
path-integral formulation which is also ineffective because of the explicit
coordinate dependence of the gauge-fixing term in the Lagrangian. One could
also try to study the spectrum of the matrix  differential operator, to find
its eigen-functions, and to use the standard expression for the  resolvent.
However, the  extension of the system for the non-zero eigenvalues leads to
unwieldy equations.  On the other hand, the Green function of the perturbation
theory must coincide with the one which solves the problem of the gauge field
interaction with the classical ``external'' current.  For this reason, we shall
compute the Green function in a most straightforward way; we shall look for the
partial solution of the inhomogeneous system using the old-fashioned method of
``variation of parameters''.  This method will immediately separate the
radiation and the longitudinal parts of the retarded propagator. All other
methods would require an additional analysis for this purpose.

Let us start the derivation of the propagator in the gauge $A^\tau=0$ by
obtaining the separate differential equations for the $\eta$-component of the
magnetic field, $\Psi= \partial_{y}A_x -\partial_{x} A_y $, the transverse
divergence of the electric field,   $\varphi= \tau (\partial_{x}
{\stackrel{\bullet} A}_x +\partial_{y}{\stackrel{\bullet} A}_y)$,  and the
$\eta$-component of the electric field, $a={\stackrel{\bullet} A}_\eta /\tau$~.
In terms of the Fourier components with respect to the spatial coordinates,
these equations read as
\begin{eqnarray}  
[\partial_{\tau}^{2}+{1\over\tau}\partial_{\tau}+ {\nu^2\over\tau^2}+
k_{\bot}^2] \Psi_{{\vec k},\nu}(\tau)= -j^\psi({\vec k},\nu,\tau)~~, 
\label{eq:E2.19}\end{eqnarray}   
\begin{eqnarray}
[\partial_{\tau}\tau\partial_{\tau}+
{\nu^2 \over \tau}] \varphi ({\vec k},\nu,\tau) -i\tau \nu k_{\bot}^{2}
a ({\vec k},\nu,\tau)
=-\partial_\tau [\tau^2 j^\varphi({\vec k},\nu,\tau) ] ~~,
\label{eq:E2.20}\end{eqnarray}             
\begin{eqnarray}
[\partial_{\tau}\tau\partial_{\tau}+\tau k_{\bot}^{2}] a({\vec k},\nu,\tau)
-{i\nu \over \tau} \varphi({\vec k},\nu,\tau) 
=-\partial_\tau [\tau^2 j^\eta ({\vec k},\nu,\tau)] ~~,              
\label{eq:E2.21}\end{eqnarray}   
where $j^\psi=\partial_{y}j_x -\partial_{x} j_y $,~
 $j^\varphi=\partial_{x}j_x +\partial_{y} j_y $.
Using  the constraint conservation, which may be explicitly integrated to
\begin{eqnarray}
 \varphi({\vec k},\nu,\tau) )+i\nu a({\vec k},\nu,\tau)- 
\tau j^\tau ({\vec k},\nu,\tau) = -\rho_0({\vec k},\nu)={\rm const}(\tau)~,  
\label{eq:E2.22}\end{eqnarray}   
one easily obtains two independent equations for $\varphi({\vec k},\nu,\tau)$
and $a({\vec k},\nu,\tau)$: 
\begin{eqnarray}
[\partial_{\tau}^{2}+{1\over \tau}\partial_{\tau}+
{\nu^2 \over \tau^2} + k_{\bot}^{2}]\varphi ({\vec k},\nu,\tau)
=k_{\bot}^2[\rho({\vec k},\nu,\tau)-\rho_0({\vec k},\nu)]-
{1\over \tau}\partial_\tau (\tau^2 j^\varphi({\vec k},\nu,\tau) )   
          \equiv f^\varphi ~~,
\label{eq:E2.23}\end{eqnarray}             
\begin{eqnarray}
[\partial_{\tau}^{2}+{1\over \tau}\partial_{\tau}+
{\nu^2 \over \tau^2} + k_{\bot}^{2}] a({\vec k},\nu,\tau)
={-i\nu\over \tau^2}[\rho({\vec k},\nu,\tau)-\rho_0({\vec k},\nu)]-
{1\over \tau}\partial_\tau (\tau^2 j^\eta ({\vec k},\nu,\tau) ) 
\equiv f^\eta ~~,              
\label{eq:E2.24}\end{eqnarray} 
The constant of integration $\rho_0({\vec k},\nu)$  has the meaning of the
arbitrary static charge density and it should be retained until Gauss' law is
explicitly imposed on the solution. In what follows, we shall not write it
explicitly, keeping in mind that it is included in the true charge density
$\rho({\vec k},\nu,\tau)$. Since wedge dynamics has a selected time moment
$\tau=0$, the constant of integration $\rho_0({\vec k},\nu)$ can be
associated with the initial data, namely, with the charge density at $\tau=0$.
As we shall see soon, a proper choice  of $\rho_0$
will be needed in order to fix the gauge $A^\tau=0$ completely.

Equations~(\ref{eq:E2.19}), (\ref{eq:E2.23}) and (\ref{eq:E2.24}) can be solved
by the method of ``variation of parameters'':     
\begin{eqnarray}
{\cal F}(\tau)= {\pi i \over 4} \int_{0}^{\tau}\tau_2d\tau_2
{\cal H}(\tau,\tau_2) f(\tau_2)~~,
\label{eq:E2.25}\end{eqnarray}     
where ${\cal F}$ stands for any one of the unknown functions in these 
equations, and $f$ for the corresponding right hand side.  The kernel
\begin{eqnarray}
{\cal H}(\tau,\tau_2)= 
H^{(1)}_{i\nu}(k_{\bot}\tau) H^{(2)}_{i\nu}(k_{\bot}\tau_2)
-H^{(2)}_{i\nu}(k_{\bot}\tau) H^{(1)}_{i\nu}(k_{\bot}\tau_2)~~.\nonumber
\end{eqnarray}
is the usual bilinear form 
which is built from the linearly-independent solutions of the
homogeneous equation. [ The Wronskian of these solutions is exactly $4/
i\pi \tau_2$. ] Taking  ${\cal F}=\Psi$,  we obtain the first equation for
the components $A_x({\vec k},\nu,\tau)$ and $A_y({\vec k},\nu,\tau)$ of
the vector  potential:
\begin{eqnarray}
\Psi({\vec k},\nu,\tau_1)\equiv
i[-k_y A_x + k_x A_y]= {i\pi\over 4} \int_{0}^{\tau_1}\tau_2d\tau_2
{\cal H}(\tau_1,\tau_2)~ i~[-k_y j^x(\tau_2) + k_x j^y(\tau_2)]~~.
\label{eq:E2.26}\end{eqnarray}                                  
In order to find the second equation for the $x$- and $y$--components 
and the equation for $A_\eta({\vec k},\nu,\tau)$, we must integrate 
twice, {\em i.e.},
\begin{eqnarray}
\Phi({\vec k},\nu,\tau_1)\equiv
i[k_x A_x + k_y A_y]= {i\pi\over 4} 
\int_{0}^{\tau_1}{d \tau' \over \tau'} \int_{0}^{\tau'} {\cal H}(\tau',\tau_2)
\tau_2 d\tau_2  ~\big[ -k_{\bot}^2 ~\rho({\vec k},\nu,\tau_2)+
{1\over \tau_2}\partial_{\tau_2} \big(\tau_{2}^{2}
j^\varphi({\vec k},\nu,\tau_2) \big)\big] ~~,
\label{eq:E2.27}\end{eqnarray}                                  
\begin{eqnarray}
A_\eta({\vec k},\nu,\tau_1) = {i\pi\over 4} 
\int_{0}^{\tau_1} \tau' d \tau' \int_{0}^{\tau'} {\cal H}(\tau',\tau_2)
\tau_2 d\tau_2 ~ \big[{i\nu\over \tau_{2}^{2}} ~\rho({\vec k},\nu,\tau_2)+
{1\over \tau_2}\partial_{\tau_2} \big(\tau_{2}^{2}
j^\eta({\vec k},\nu,\tau_2) \big)\big] ~~.
\label{eq:E2.28}\end{eqnarray}              
The integration over $\tau_2$ recovers the electric fields at the
moment $\tau'$, whilst the integration over $\tau'$ gives the vector
potential at the moment $\tau_1$. It is convenient to start with the second
of these integrations
which has the limits  $\tau_2 <\tau'<\tau_1$. Let us consider the main
line of the calculations in detail,
using the $\eta$-component as an example.
The first integration follows the formula (\ref{eq:A2.1}),
\begin{eqnarray}   
k_{\bot}^{\mu +1}\int_{\tau_2}^{\tau_1} (\tau')^\mu 
H^{(j)}_{i\nu}(k_{\bot}\tau') d \tau' = 
R^{(j)}_{\mu,i\nu}(k_{\bot}\tau_1)-R^{(j)}_{\mu,i\nu}(k_{\bot}\tau_2)~~,
\label{eq:E2.29}\end{eqnarray}    
and  the terms emerging from the lower limit $\tau_2$ can be conveniently 
transformed according to the relation (see Appendix~2),
\begin{eqnarray}   
R^{(1)}_{\mu,i\nu}(k_{\bot}\tau_2)H^{(2)}_{i\nu}(k_{\bot}\tau_2) -
R^{(2)}_{\mu,i\nu}(k_{\bot}\tau_2)H^{(1)}_{i\nu}(k_{\bot}\tau_2)     
={4\over i\pi}s_{\mu,i\nu}(k_{\bot}\tau_2)~.
\label{eq:E2.30}\end{eqnarray}      
As a result, one obtains, {\em e.g.}, the following formula for 
 $A_\eta({\vec k},\nu,\tau_1)$:
\begin{eqnarray}
A_\eta({\vec k},\nu,\tau_1) = {i\pi\over 4 k_{\bot}^2}
\int_{0}^{\tau_1}\tau_2 d\tau_2
[R^{(1)}_{1,i\nu}(k_{\bot}\tau_1)H^{(2)}_{i\nu}(k_{\bot}\tau_2) -
R^{(2)}_{1,i\nu}(k_{\bot}\tau_1)H^{(1)}_{i\nu}(k_{\bot}\tau_2) 
- {4\over i\pi}s_{1,i\nu}(k_{\bot}\tau_2)] \nonumber \\  
\times \big[{i\nu\over \tau_{2}^{2}} \rho({\vec k},\nu,\tau_2)+
{1\over \tau_2}\partial_{\tau_2} \big(\tau_{2}^{2}
j^\eta({\vec k},\nu,\tau_2) \big) \big]~~.
\label{eq:E2.31}\end{eqnarray}          
In order to eliminate the charge density $\rho$ from the integrand and to
separate the transverse and the longitudinal parts of the propagator, all
the terms of this formula should be integrated by parts, explicitly
accounting for the charge conservation, which reads as
\begin{eqnarray}
i\tau [k_x j^x({\vec k},\nu,\tau) + k_x j^y({\vec k},\nu,\tau)
+\nu j^\eta({\vec k},\nu,\tau)]+\partial_\tau \rho({\vec k},\nu,\tau)=0~~.
\label{eq:E2.32}\end{eqnarray}  
We have in sequence:
\begin{eqnarray}   
i\nu\int_{0}^{\tau_1} {d\tau_2\over\tau_2} \rho(\tau_2)
H^{(j)}_{i\nu}(k_{\bot}\tau_2) = i\nu\int_{0}^{\tau_1} 
{d R^{(j)}_{-1,i\nu}(k_{\bot}\tau_2) \over d\tau_2} 
\rho(\tau_2)d\tau_2 \nonumber\\ 
=i\nu R^{(j)}_{-1,i\nu}(k_{\bot}\tau_1) \rho(\tau_1)-
\nu \int_{0}^{\tau_1} \tau_2 d\tau_2 R^{(j)}_{-1,i\nu}(k_{\bot}\tau_2)      
[k_x j^x(\tau_2) + k_y j^y(\tau_2)+\nu j^\eta(\tau_2)]~~,
\label{eq:E2.33}\end{eqnarray}        
\begin{eqnarray}   
i\nu\int_{0}^{\tau_1} d\tau_2 
H^{(j)}_{i\nu}(k_{\bot}\tau_2)\partial_{\tau_2} (\tau_{2}^{2}
j^\eta(\tau_2) ) 
= \tau_{1}^{2} j^\eta(\tau_1)H^{(j)}_{i\nu}(k_{\bot}\tau_1)  
+ \int_{0}^{\tau_1} \tau_2 d\tau_2 [R^{(j)}_{1,i\nu}(k_{\bot}\tau_2)      
 +\nu^2 R^{(j)}_{-1,i\nu}(k_{\bot}\tau_2)]j^\eta(\tau_2).
\label{eq:E2.34}\end{eqnarray}        
In a similar way we have,
\begin{eqnarray}   
i\nu\int_{0}^{\tau_1} {d\tau_2\over\tau_2} \rho(\tau_2)
s_{1,i\nu}(k_{\bot}\tau_2) = i\nu\int_{0}^{\tau_1}
{dQ_{-1,i\nu}(k_{\bot}\tau_2)\over d\tau_2} \rho(\tau_2)d\tau_2 \nonumber\\ 
=i\nu Q_{-1,i\nu}(k_{\bot}\tau_1) \rho(\tau_1)-
\nu \int_{0}^{\tau_1} \!\! \tau_2 d\tau_2 Q_{-1,i\nu}(k_{\bot}\tau_2)      
[k_x j^x(\tau_2) + k_y j^y(\tau_2)+\nu j^\eta(\tau_2)]~~,
\label{eq:E2.35}\end{eqnarray}       
\begin{eqnarray}   
\int_{0}^{\tau_1} d\tau_2 s_{1,i\nu}(k_{\bot}\tau_2)
\partial_{\tau_2} (\tau_{2}^{2}j^\eta(\tau_2) ) 
= \tau_{1}^{2}
j^\eta(\tau_1)s_{1,i\nu}(k_{\bot}\tau_1)  
+ \nu^2 \int_{0}^{\tau_1} \tau_2 d\tau_2 [Q_{-1,i\nu}(k_{\bot}\tau_2)      
-Q_{1,i\nu}(k_{\bot}\tau_2)]j^\eta(\tau_2)~~.
\label{eq:E2.36}\end{eqnarray}  
Assembling these pieces together and repeating the same calculations
for the function $\Phi$ one obtains three different terms which contribute
to the field $A$ produced by the current $j$;~~
$A=A^{(tr)}+A^{(L)}+A^{(inst)}$.

The transverse field $A^{(tr)}$ is defined by the integral terms in 
the R.H.S. of Eqs.~(\ref{eq:E2.33}) and (\ref{eq:E2.34}). It can be
conveniently written down in the following form,
\begin{eqnarray}
A^{(tr)}_{l}(x_1)=\int d^4 x_2 \theta(\tau_1-\tau_2)
\Delta^{(tr)}_{lm}(x_1,x_2) j^m(x_2)~~.
\label{eq:E2.37}\end{eqnarray}     
where                             
\begin{eqnarray}
\Delta^{(tr)}_{lm}(x,y)= -i\int_{-\infty}^{\infty} 
d\nu\int d^2{\vec k}\sum_{\lambda=1,2}
[V^{(\lambda)}_{\nu {\vec k};l}(x) V^{(\lambda)\ast }_{\nu {\vec k};m}(y)-
V^{(\lambda)\ast}_{\nu {\vec k};l}(x) V^{(\lambda) }_{\nu {\vec k};m}(y)]~~,
\label{eq:E2.38}\end{eqnarray}   
which can be easily recognized as the Riemann function of the original
homogeneous hyperbolic system. The Riemann function solves the boundary
value problem for the evolution of the  free radiation field.  It is
obtained immediately in the form of the bilinear expansion over the full
set of solutions  (\ref{eq:E2.15}) of the homogeneous system. In fact,
this is a sole evidence that  $\Delta^{(tr)}$ may be associated with the 
transverse part of the propagator. Then the remaining part is the 
propagator (response function) for the longitudinal field.

The dynamical  longitudinal field $A^{(L)}$ originates from the integral 
terms in  the R.H.S. of Eqs.~(\ref{eq:E2.35}) and (\ref{eq:E2.36}): 
\begin{eqnarray}
A^{(L)}_{l}(\tau_1,\eta_1,{\vec r}_1)=\int_{0}^{\tau_1} \tau_2 d\tau_2 
\int  d\eta_2 d^2{\vec r}_2
\Delta^{(L)}_{lm}(\tau_2;\eta_1-\eta_2,{\vec r}_1-{\vec r}_2) 
j^m(\tau_2,\eta_2,{\vec r}_2)~~.     
\label{eq:E2.39}\end{eqnarray}         
The kernel of this representation,        
\begin{eqnarray}
\Delta^{(L)}_{lm}(\tau_2;\eta_1-\eta_2,{\vec r}_1-{\vec r}_2) 
= \int {d\nu d^2{\vec k} \over (2\pi)^3 k_{\bot}^2}
 \left[ \begin{array}{c} 
                 k_r  \\ 
                 \nu  \end{array} \right]_l 
\left[ \begin{array}{c} 
                 k_s  Q_{-1,i\nu}(k_{\bot}\tau_2) \\ 
                 \nu Q_{1,i\nu}(k_{\bot}\tau_2)
                                            \end{array} \right]_m 
              e^{i\nu(\eta_1-\eta_2) +i{\vec k}({\vec r}_1-{\vec r}_2)}~~,     
\label{eq:E2.40}\end{eqnarray}           
does not allow for the bilinear expansion with two temporal arguments,
and, as we shall see in a while, the retarded character of the integration
in Eq.~(\ref{eq:E2.39}) is not sensitive to the light cone
boundaries. In fact, the
electric field $E_{l}^{(L)}={\stackrel{\bullet} A}_{l}^{(L)}$ is simultaneous with the 
current $j^m$ .

The  instantaneous part of the solution comes from the 
boundary terms in Eqs.~(\ref{eq:E2.33})--(\ref{eq:E2.36}) which were
generated via  integration by parts.  It depends on a single time
variable $\tau_1$. Using two functional relations, (\ref{eq:E2.30}) and
\begin{eqnarray}   
R^{(1)}_{1,i\nu}(x)R^{(2)}_{-1,i\nu}(x)   -
R^{(2)}_{1,i\nu}(x)R^{(1)}_{-1,i\nu}(x)       
=-{4\over i\pi}{x\over\nu^2}{d s_{1,i\nu}(x)\over d x}=
-{4\over i\pi}[Q_{1,i\nu}(x)-Q_{-1,i\nu}(x)] ~~,
\label{eq:E2.41}\end{eqnarray}      
( see Appendix~2 ) its Fourier transform can be presented in the form 
\begin{eqnarray} 
A^{(inst)}_{l}({\vec k},\nu;\tau_1)=
{ \rho({\vec k},\nu,\tau_1) \over (2\pi)^3 i k_{\bot}^2}
 \left[ \begin{array}{c} 
                 k_r  Q_{-1,i\nu}(k_{\bot}\tau_1) \\ 
                 \nu Q_{1,i\nu}(k_{\bot}\tau_1)
                                            \end{array} \right]_l~~,     
\label{eq:E2.42}\end{eqnarray}                     
which leads to the Poisson-type integral,
\begin{eqnarray} 
A^{(inst)}_{m}(\tau_1,\eta_1,{\vec r}_1) = \int d{\vec r}_2 d\eta_2 
{\cal K}_m(\tau_1;{\vec r}_1-{\vec r}_2,\eta_1 -\eta_2) 
\rho(\tau_1,{\vec r}_2,\eta_2)~,
\label{eq:E2.43}\end{eqnarray}    
with the {\em instantaneous} kernel,
\begin{eqnarray}   
{\cal K}_m(\tau;{\vec r},\eta)= \int {d\nu d{\vec k}\over (2\pi)^3}
{e^{i\nu\eta +i{\vec k}{\vec r}}\over i k_{\bot}^2}
\left[ \begin{array}{c} 
                 k_r   Q_{-1,i\nu}(k_{\bot}\tau) \\ 
                 \nu  Q_{1,i\nu}(k_{\bot}\tau)
                                            \end{array} \right]_m~~.
\label{eq:E2.44}\end{eqnarray} 
The potential $A^{(inst)}$ given by Eq.~(\ref{eq:E2.42}) coincides with the
potential $V^{(3)}$ of Eq.~(\ref{eq:E2.16}) of the $\tau$-static source. 
Therefore, this term represents the instantaneous distribution of the potential
at the moment $\tau_1$, corresponding to the charge density taken at the same
moment.  Next, we have to recall that the charge density   $\rho({\vec
k},\nu,\tau_1) $ in Eq.~(\ref{eq:E2.42}) still includes an arbitrary constant
$\rho_0({\vec k},\nu)$, which may be interpreted as the charge density at
$\tau=0$. This constant has appeared because we used only the conservation 
(\ref{eq:E2.22}) of the constraint (which is the consequence of the equations
of motion) and did not used the Gauss law explicitly. Now we can see that
imposing the constraint indeed affects only the static potential of the charge
distribution and puts it in agreement with Gauss law.  If the initial data
allow one to put  $\rho_0 =\rho(\tau=0)=0$, then it immediately solves two
problems. First, the conservation of the constraint just duplicates the Gauss
law, and the latter can be used to remove the unphysical degrees of freedom
without any reservations. Eventually, it allows to fix the gauge $A^\tau =0$
completely (see Sec.~\ref{subsec:SB30}).  
Second, it becomes possible to eliminate the charge
density $\rho$  completely, and to replace it by the components  $j^n$ of the
current.  Replacement follows an evident prescription, 
\begin{eqnarray}
\rho(\tau_1,\eta_2,{\vec r}_2)= \int_{0}^{\tau_1}d\tau_2 {\partial\rho
\over\partial\tau_2} = -i \int_{0}^{\tau_1}\tau_2 d\tau_2
[k_s j^s(\tau_2,\eta_2,{\vec r}_2)+
\nu j^\eta(\tau_2,\eta_2,{\vec r}_2)]~~, \nonumber   
\end{eqnarray} 
and leads to the standard form of the $A^{(inst)}$ representation (an
artificial contribution of any $\rho_0$ would correspond to the  recognizable
static pattern in the longitudinal part of the propagator and is easily
handled), 
\begin{eqnarray}
A^{(inst)}_{l}(\tau_1,\eta_1,{\vec r}_1)=\int_{0}^{\tau_1} \tau_2 d\tau_2 
\int  d\eta_2 d^2{\vec r}_2
\Delta^{(inst)}_{lm}(\tau_1;\eta_1-\eta_2,{\vec r}_1-{\vec r}_2) 
j^m(\tau_2,\eta_2,{\vec r}_2)~,     
\label{eq:E2.43a}\end{eqnarray}    
with the kernel given by the formula,
\begin{eqnarray}
\Delta^{(inst)}_{lm}(\tau_1;\eta_1-\eta_2,{\vec r}_1-{\vec r}_2 )
= - \int {d\nu d^2{\vec k} \over (2\pi)^3 k_{\bot}^2}
\left[ \begin{array}{c} 
                 k_r  Q_{-1,i\nu}(k_{\bot}\tau_1) \\ 
                 \nu Q_{1,i\nu}(k_{\bot}\tau_1)
                                            \end{array} \right]_l 
 \left[ \begin{array}{c} 
                 k_s  \\ 
                 \nu  \end{array} \right]_m    
               e^{i\nu(\eta_1-\eta_2) +i{\vec k}({\vec r}_1-{\vec r}_2)}~~.     
\label{eq:E2.44a}\end{eqnarray}              

Two parts of the propagator, given by equations 
(\ref{eq:E2.40}) and (\ref{eq:E2.44a}) can be combined in one elegant
formula for the propagator of the field $A^{(long)} = A^{(L)}+A^{(inst)}$,
\begin{eqnarray}
\Delta^{(long)}_{lm}(\tau_1;\eta_1-\eta_2,{\vec r}_1-{\vec r}_2 )
=\hspace{12cm}     \label{eq:E2.44b}\\
=  \int {d\nu d^2{\vec k} \over (2\pi)^3 k_{\bot}^2}
\left[ \begin{array}{cc} 
 k_rk_s [Q_{-1,i\nu}(k_{\bot}\tau_2)- Q_{-1,i\nu}(k_{\bot}\tau_1)] &
 k_r \nu [ Q_{1,i\nu}(k_{\bot}\tau_2) -Q_{-1,i\nu}(k_{\bot}\tau_1)]\\
 \nu k_s  [ Q_{-1,i\nu}(k_{\bot}\tau_2) -Q_{1,i\nu}(k_{\bot}\tau_1)] &
 \nu^2  [ Q_{1,i\nu}(k_{\bot}\tau_2) -Q_{1,i\nu}(k_{\bot}\tau_1)]
                                            \end{array} \right]_{lm} 
 e^{i\nu(\eta_1-\eta_2) +i{\vec k}({\vec r}_1-{\vec r}_2)}~.\nonumber     
\end{eqnarray}  
This expression will be used for practical calculation of quark self-energy in
paper [IV], where it will be transformed into the mixed representation. The
coordinate form of $A^{(L)}$ that reveals its causal properties, is
analyzed in Sec.~\ref{sec:SN5}

Eqs. (\ref{eq:E2.37})--(\ref{eq:E2.40}) and (\ref{eq:E2.43a}), 
(\ref{eq:E2.44a}) present the propagator in a split form. Different
constituents of this form are as a preliminary identified as the transverse,  the
longitudinal and  the instantaneous parts of the propagator. It would be useful
to learn if the same kind of splitting is possible for the current itself. An
affirmative answer (as in the cases of the Coulomb and radiation gauges) would
be  helpful for the design of the perturbation theory.  To answer this
question, one should substitute the different pieces of the solution  into the
left hand side of the original system of differential equations. This leads to
the following expressions for the Fourier components of the three currents:
\begin{eqnarray}   
\tau j^{m}_{(tr)}({\vec k},\nu;\tau)=\tau j^{m}({\vec k},\nu;\tau)+
 {1\over ik_{\bot}^{2}}     \left[ \begin{array}{c} 
    k_r   s_{1,i\nu}(k_{\bot}\tau) \\ 
    \nu k_{\bot}^{2}  s_{-1,i\nu}(k_{\bot}\tau)  
    \end{array} \right]^m {\partial\rho\over\partial\tau} -
{\nu\over k_{\bot}^{2}} {\partial\over\partial\tau}
               \bigg( {\stackrel{\bullet} s}_{-1,i\nu}(k_{\bot}\tau)
\left[ \begin{array}{c} 
                 k_r  \tau^3 j^\eta  \\ 
                 -\tau (k_xj^x+k_yj^y)
                             \end{array} \right]^m \bigg)~~,
\label{eq:E2.45}\end{eqnarray} 
\begin{eqnarray}   
\tau j^{m}_{(L)}({\vec k},\nu;\tau)={1\over k_{\bot}^{2}} 
 {\partial\over\partial\tau} \bigg(
\left[ \begin{array}{c} 
                 k_r\tau^2 \\ \nu \end{array} \right]^m 
[Q_{-1,i\nu}(k_{\bot}\tau)(k_xj^x+k_yj^y) 
+Q_{1,i\nu}(k_{\bot}\tau) j^\eta ]\bigg)~~,
\label{eq:E2.45a}\end{eqnarray} 
\begin{eqnarray}   
\tau j^{m}_{(inst)}({\vec k},\nu;\tau)= {-1\over ik_{\bot}^{2}}
        {\partial\over\partial\tau} 
    \bigg( \left[ \begin{array}{c} 
                 k_r   \tau Q_{-1,i\nu}(k_{\bot}\tau) \\ 
                 \nu  \tau^{-1} Q_{1,i\nu}(k_{\bot}\tau)
                  \end{array} \right]^m
{\partial\rho\over\partial\tau}  \bigg) -{1\over ik_{\bot}^{2}}\ 
\left[ \begin{array}{c} 
                 k_r   s_{1,i\nu}(k_{\bot}\tau) \\ 
                 \nu k_{\bot}^{2}  s_{-1,i\nu}(k_{\bot}\tau)
                                            \end{array} \right]^m 
                       {\partial\rho\over\partial\tau}~~, 
\label{eq:E2.46}\end{eqnarray} 
Provided that  the current is conserved, these three currents, added together,
give the full current on the right hand side of the system. Therefore, the
solution is correct. However, none of these three currents  carries any
signature of being longitudinal or transversal in the usual sense. None of
them  has zero divergence since the operator of the divergence does not commute
with the differential operator of the system. No desired simplification is
possible in our case.

In fact,  the above splitting of the potential has no real physical
meaning. To see it explicitly, let us find the divergence of the electric
field, ${\rm div}{\bf E}= \partial_m{\cal E}^m $ [again, for brevity, in the
Fourier representation]:
\begin{eqnarray}  
{\rm div}{\bf E}^{(tr)}({\vec k},\nu;\tau) 
=i[Q_{-1,i\nu}(k_{\bot}\tau)-Q_{1,i\nu}(k_{\bot}\tau)]
\big(\nu\tau^2 j^\eta - {\nu^2\over k_{\bot}^{2}}(k_xj^x+k_yj^y)\big)~~, 
\label{eq:E2.46A}\end{eqnarray}  
\begin{eqnarray}  
{\rm div}{\bf E}^{(L)}({\vec k},\nu;\tau)
=i \big( \tau^2+{\nu^2\over k_{\bot}^{2}}\big)
[(k_xj^x+k_yj^y) Q_{-1,i\nu}(k_{\bot}\tau)
+\nu j^\eta Q_{1,i\nu}(k_{\bot}\tau)]~~,
\label{eq:E2.46B}\end{eqnarray}  
\begin{eqnarray}  
{\rm div}{\bf E}^{(inst)}({\vec k},\nu;\tau)=
\rho({\vec k},\nu;\tau) -i \big( \tau^2 Q_{-1,i\nu}(k_{\bot}\tau) 
-{\nu^2\over k_{\bot}^{2}} Q_{1,i\nu}(k_{\bot}\tau)\big)
[(k_xj^x+k_yj^y) +\nu j^\eta]~~.
\label{eq:E2.46C}\end{eqnarray}  
Only the divergence of the true retarded component of the field
${\bf E}^{(tr)}$ turns out to be zero. The term which prevents the  
${\rm div}{\bf E}^{(tr)}$ from being zero is due to the non-symmetry of the
propagator,  $\Delta^{\eta r}\neq\Delta^{r\eta}$. It appears when the
$\theta$-function in Eq.~(\ref{eq:E2.37}) is  differentiated with respect to
Hamiltonian  time $\tau$. This term is vital for obtaining  the expression
that obeys the Gauss constraint, ${\rm div}{\bf E}({\vec k},\nu;\tau)=
\rho({\vec k},\nu;\tau)$.  

The known examples, when the transverse and  the longitudinal fields are
separated at the level of the equations of motion, are  related to a narrow
class of homogeneous gauges.  The impossibility of a universal separation of
the transverse and longitudinal currents thus appears to be a rule rather than
exception. It reflects a general principle; the radiation field created at some
time interval has the preceding and the subsequent configurations of the
longitudinal field as the boundary condition. The dynamics of the longitudinal
field falls out of any scattering problem in its $S$-matrix formulation.
However, this dynamics is, in fact, a subject of the QCD evolution in the
inelastic high-energy processes.

\subsection{Initial data problem in the gauge $A^{\tau}=0$} 
\label{subsec:SB25}

We obtained the expression for the (retarded) propagator as the response 
function between the ``external'' current and the potential of the gauge 
field. We must also verify that the same propagator solves the Cauchy
problem for the gauge field. This can be easily done by presenting
the initial data at the surface $\tau=\tau_0$ in the form of the 
source density at the hyper-surface $\tau=\tau_0$, {\em i.e.},
\begin{eqnarray} 
\sqrt{-{\rm g}}J^n(\tau_2)=\sqrt{-{\rm g}(\tau_0)}{\rm g}^{nm}(\tau_0)
[\delta'(\tau_2-\tau_0){\bar A}_m({\vec r},\eta)+
\delta(\tau_2-\tau_0){\bar A'}_m({\vec r},\eta) ]~,
\label{eq:E2.47}\end{eqnarray}
where ${\bar A}_m({\vec r},\eta)$ and ${\bar A'}_m({\vec r},\eta)$ are
the initial data for the potential and its normal derivative on the
hyper-surface $\tau = \tau_0$.
Usually, it is assumed that the external current vanishes for $\tau<\tau_0$.  
Substituting this source into the Eqs.~(\ref{eq:E2.37}), (\ref{eq:E2.39}) 
and (\ref{eq:E2.43}), and taking the limit of $\tau_1\rightarrow\tau_0$,
we may verify that the standard prescription for the solution of the
initial data problem,
\begin{eqnarray}
A_{l}(x_1)=\int_{(\tau_2=\tau_0)} d^2{\vec r}_2~d\eta_2~\Delta_{lm}(x_1,x_2) 
{\stackrel{\leftrightarrow}{\partial\over\partial\tau_2}}
A^m(x_2)~~,
\label{eq:E2.48}\end{eqnarray}     
holds with the same propagator $\Delta_{lm}(x_1,x_2) $ that was used to
solve the emission problem. For example, in the limit of
$\tau\rightarrow\tau_0$, the $\eta$-component of the vector 
potential is a sum of three terms, 
\begin{eqnarray} 
A^{(tr)}_{\eta}(\tau_0+0)={i\pi\over 4k_{\bot}^{2}}~\big\{~ 
[R^{(2)}_{1,i\nu}(k_{\bot}\tau_0)H^{(1)}_{i\nu}(k_{\bot}\tau_0)-
R^{(1)}_{1,i\nu}(k_{\bot}\tau_0)H^{(2)}_{i\nu}(k_{\bot}\tau_0)]
[\nu {\bar A}_\phi -k_{\bot}^{2}{\bar A}_\eta] \nonumber \\   
-\tau_0\nu 
[R^{(2)}_{1,i\nu}(k_{\bot}\tau_0) R^{(1)}_{-1,i\nu}(k_{\bot}\tau_0)-
R^{(1)}_{1,i\nu}(k_{\bot}\tau_0) R^{(2)}_{-1,i\nu}(k_{\bot}\tau_0)]
{\bar A'}_\phi ~\big\} ~~,
\label{eq:E2.49}\end{eqnarray}      
\begin{eqnarray} 
A^{(L)}_{\eta}(\tau_0+0)={- \nu\over k_{\bot}^{2}}
\big\{~-s_{1,i\nu}(k_{\bot}\tau_0)
{\bar A}_\phi + \tau_0 Q_{-1,i\nu}(k_{\bot}\tau_0){\bar A'}_\phi   
-{\nu k_{\bot}^{2}\over\tau_0} s_{-1,i\nu}(k_{\bot}\tau_0){\bar A}_\eta
+{\nu \over\tau_0} Q_{1,i\nu}(k_{\bot}\tau_0){\bar A'}_\eta~\big\}~~,
\label{eq:E2.50}\end{eqnarray}              
\begin{eqnarray} 
A^{(inst)}_{\eta}(\tau_0+0)= {\nu\over k_{\bot}^{2}}
Q_{1,i\nu}(k_{\bot}\tau_0) [\tau_0{\bar A'}_\phi 
+{\nu \over\tau_0}{\bar A'}_\eta ]~~,
\label{eq:E2.51}\end{eqnarray}
where we have denoted: ${\bar A}_\phi = k_x{\bar A}_x +k_y{\bar A}_y $.    
 Eq.~(\ref{eq:E2.51}) follows from Eq.~(\ref{eq:E2.43}) and takes care 
of the consistency between the charge density at the moment $\tau_0$ and the
initial data for the gauge field. Using relations (\ref{eq:E2.30}) and
(\ref{eq:E2.42}) and adding up Eqs.~(\ref{eq:E2.49})--(\ref{eq:E2.51})
we come to a desired identity, $A_{\eta}(\tau_0+0)= {\bar A}_\eta $.

When the initial data ${\bar A}_m({\vec r},\eta)$   and  ${\bar A'}_m
({\vec r},\eta)$ correspond to the free radiation field, then only the part
of the full propagator, $\Delta^{(tr)}_{lm}(x_1,x_2)$, ``works'' here, and
only Eq.~(\ref{eq:E2.49}) may be retained. The other two equations acquire
the status of being constraints  imposed on the initial data. Since
the current is absent, we have $A^{(L)} =0$ on the left hand side of the
Eqs.~(\ref{eq:E2.50}). Then the right hand side confirms that the
kernel  ${\cal K}$ is orthogonal to the free radiation field modes.
Since the charge density $\rho$ vanishes, we have 
$A^{(inst)} =0$, which is equivalent to Gauss' law for the free gauge
field. The two transverse modes already obey these constraints.
This fact provides a reliable footing for the canonical
quantization of the free field in the gauge $A^\tau=0$. Indeed, the Riemann
function coincides with the commutator of the free gauge field. It can be
found via its bilinear decomposition over the physical modes. Thus, one
can avoid technical problems of inverting the constraint equations (see
Sec.~\ref{sec:SN3}.) The longitudinal part of the propagator will be
studied in details in Sec.~\ref{sec:SN5}. 

\subsection{Gluon vertices in the gauge $A^{\tau}=0$} 
\label{subsec:SB24} 
 
The terms proportional to the first and the second powers of the coupling
constant in the classical wave equations may be viewed as the external
current and allow one to define the explicit form of the 3- and 4-gluon
vertices. One should start from the solution of the Maxwell equations,
\begin{eqnarray}
A^{a'}_{k'}(z_1)=\int d^4 x
\Delta^{a'a}_{k'k}(z_1,x) \sqrt{-{\rm g}(x)} {\cal J}^{k}_{a}(x)~,     
\label{eq:E2.52}\end{eqnarray}      
with the color current of the form
\begin{eqnarray}
\sqrt{-{\rm g}(x)} {\cal J}^{k}_{a}(x)=-gf_{abc}\sqrt{-{\rm g}(x)}
{\rm g}^{kn}(x){\rm g}^{ml}(x) [\partial_m(A^{b}_{l}(x)A^{c}_{n}(x))+
A^{b}_{l}(x) \partial_m A^{c}_{n}(x) + 
A^{b}_{m}(x)\partial_n A^{c}_{l}(x) ] \nonumber \\
-g^2 \sqrt{-{\rm g}(x)} f_{abc}f_{cdh} {\rm g}^{kn}(x){\rm g}^{ml}(x)
A^{b}_{l}(x)A^{d}_{m}(x)A^{h}_{n}(x)~~.     
\label{eq:E2.53}\end{eqnarray} 
In perturbation calculations, every field $A(x)$ in the RHS of this
expression is a part of some correlator $\Delta(x,z_N)$. The components of
the metric depend only on the time $\tau$ while the derivatives affect only
the spatial directions $u^n=({\vec r},\eta)$. Moreover, in these
directions, all the gluon correlators depend only on the differences
of the coordinates and can be rewritten in terms of their spatial Fourier
components. After symmetrization over the outer arguments $z_N$, one
immediately obtains,
\begin{eqnarray}
V^{kln}_{abc}(p_1,p_2,p_3;\tau)=-i\tau f_{abc}~\delta(p_1+p_2+p_3)~ 
[{\rm g}^{ln}(p_2-p_3)^k+
{\rm g}^{nk}(p_3-p_1)^l+{\rm g}^{kl}(p_1-p_2)^n]~~,     
\label{eq:E2.54}\end{eqnarray}        
where $p^n={\rm g}^{nk}p_k$, and the components of the momentum in the 
curvilinear coordinates are equal to $p_k=(p_x,p_y,\nu)$.
The four-gluon vertex has no derivatives and is the same as usually.

\section{Quantization}   
\label{sec:SN3} 

The second quantization of the field has several practical goals. 
We would like to have an expansion of the  operator of the free
gluon field like
\begin{equation} 
A_i (x)=\sum_{\lambda=1,2} \int d^2 {\vec k} d\nu [ c_{\lambda}(\nu,{\vec k})
V^{(\lambda)}_{\nu {\vec k};i}(x) + c^{\dag}_{\lambda}(\nu,{\vec k})  
V^{(\lambda)\ast }_{\nu {\vec k};i}(x)]~~,              
\label{eq:E3.1}\end{equation}        
with the creation and annihilation operators which obey the commutation
relations
\begin{eqnarray}
[c_{\lambda}(\nu,{\vec k}),c^{\dag}_{\lambda'}(\nu',{\vec k}')] 
=\delta_{\lambda\lambda'}\delta(\nu-\nu')\delta({\vec k}-{\vec k}'),\;\;\;
[c_{\lambda}(\nu,{\vec k}),c_{\lambda'}(\nu',{\vec k}')]= 
[c^{\dag}_{\lambda}(\nu,{\vec k}),
c^{\dag}_{\lambda'}(\nu',{\vec k}')]=0~~. 
\label{eq:E3.2}\end{eqnarray} 
Once obtained, the commutation relations (\ref{eq:E3.2}) allow one to find
various correlators of the free gluon field as the averages of the binary
operator products over the state of the perturbative vacuum and express
them  via the solutions $V^{(1)}_{\nu {\vec k};i}(x)$ and  
$V^{(2)}_{\nu{\vec k};i}(x)$. For example, the  Wightman functions,
\begin{eqnarray}
i\Delta_{10,ij}(x,y)= \langle 0|  A_i(x) A_j(y) | 0 \rangle =
\sum_{\lambda=1,2}\int d\nu d^2{\vec k}
V^{(\lambda)}_{\nu {\vec k};i}(x) V^{(\lambda)\ast }_{\nu {\vec k};i}(y)
=i\Delta_{01,ji}(y,x)~~,
\label{eq:E3.3}\end{eqnarray} 
serve as the projectors onto the space of the on-mass-shell  gluons and should
be known explicitly in order to have a good  definition of the production rate
of  gluons in the final states. The Fock creation and annihilation operators
are also needed in order to define the occupation numbers and to introduce the
gluon distributions into various field correlators defined as the averages over
an ensemble. With these two Wightman functions at hand, one immediately obtains
the expression for the commutator of the free field operators,
\begin{eqnarray}
\Delta_{0,ij}(x,y)= -i \langle 0|[  A_i(x), A_j(y)] | 0 \rangle =
  \Delta_{10,ij}(x,y)-\Delta_{01,ij}(x,y)~,
\label{eq:E3.3a}\end{eqnarray}    
which should coincide with the Riemann function of the homogeneous field
equations.  The program of second quantization does not reveal any
technical problems if we give preference to the holomorphic quantization
which is based on commutation relations  (\ref{eq:E3.2}) for the Fock
operators. However, if we prefer to start with the canonical commutation
relations for the field coordinates and momenta, then one should {\em
postulate} them and derive (\ref{eq:E3.2}) as the consequence.
 
The way to obtain  the canonical commutation relations in cases of the
scalar  and the spinor fields  is quite straightforward. For the vector
gauge  field, we meet a well-known problem,
{\em viz.}, an excess of the number of the
components of the vector field  over the number of the physical degrees of
freedom. For example, in the  so-called radiation gauge, $A^0=0$ and
${\rm div} {\bf A}=0 $, we write the canonical commutation relations 
in the following form \cite{Bjorken},
\begin{eqnarray}       
\big[ A_{i} ({\bf x},t), E_{j} ({\bf y},t) \big]=
\delta_{ij}^{tr} ({\bf x}-{\bf y})
=\int {d^3 {\bf k}\over (2\pi )^3 } \bigg( \delta_{ij}-
{ k_i k_j \over {\bf k}^2 } \bigg) e^{-ik(x-y)} ,  \nonumber    \\
\big[ A_{i} ({\bf x},t), A_{j} ({\bf y},t) \big]  = 
\big[ E_{i} ({\bf x},t), E_{j} ({\bf y},t) \big]=0~, 
\label{eq:E3.4}\end{eqnarray} 
thus eliminating  the longitudinally polarized photons from the dynamical
degrees of freedom. The function $\delta_{ij}^{tr}$  plays a role as the
unit operator in the space of the physical states. Here, $i,j=1,2,3$ and 
the number of relations postulated by equations (\ref{eq:E3.4}) apparently
exceeds the actual number required by the count of the independent degrees
of freedom, $\lambda=1,2$, of the free gauge field. The Fourier transform
of the function $\delta_{ij}^{tr}$ is easily guessed because the basis  
of the plane-wave  solutions is very simple \cite{Bjorken}, and it can be
obtained rigorously by solving the system of the constraint equations
\cite{Weinberg,Tyutin}. A similar guess or procedure in our  case is not
so obvious. We have the  gauge condition $A^\tau=0$ as the primary
constraint and Gauss' law as the secondary one. The latter can be
resolved in a way which allows one to exclude the $\eta$-components of the
potential and the electric field  from the set of independent canonical
variables. Thus, only $x$- and $y$-components are subject to the canonical
commutation relations. To resolve the constraints, one  needs the
integral operators with the kernels built from the solutions of the
Maxwell equations in the gauge $A^\tau =0$. Therefore we shall proceed in
two steps. In section ~\ref{subsec:SB31}, we shall sketch the results for
the Wightman functions  (\ref{eq:E3.3}). These, will be used for the
explicit calculation of the free field commutator (\ref{eq:E3.3a}) in
section ~\ref{subsec:SB32} and for the study of its causal behavior.

\subsection{Fixing of the gauge $A^{\tau}=0$} 
\label{subsec:SB30} 

Only the independent components of transverse fields are the subject for
quantization. In order to eliminate the extraneous degrees of freedom we have
to incorporate the Gauss law. In Sec.~\ref{subsec:SB23}, the latter was shown
to be a consequence of the equations of motion only when the evolution begins
with the zero charge density at $\tau=0$. Are these initial data physical or
do they mean that the QCD evolution begins from nothing? Addressing the hadron
collisions, the question can be asked more specifically: do we really need any
resolved {\em ad hoc} color charges (dipoles, quadrupoles, etc.) to initiate
the color interaction?  The answer is negative for three reasons. First, as it
has been demonstrated in paper [II], at $\tau\to 0$ the states of wedge
dynamics are widely spread along the null-planes with zero density at $\tau=
0$. Second, the nucleus which fluctuate into the expanded state of wedge
dynamics is colorless even locally. Third, as it will be demonstrated in paper
[IV], the interactions which are the strongest at very early times, are
magneto-static by their nature. The collision of two nuclei is more likely to
begin with the magnetic interaction of the color currents in the locally
color-neutral system than with the electric interaction of  color charges.

In Abelian case considered here the gauge transformation is
\begin{eqnarray}
A'_\mu (x)= A_\mu (x) +\partial_\mu \chi(x)~.
\label{eq:E3.01} \end{eqnarray} 
Since we have $A_\tau(x)=0$, we must also have $\chi(x)=\chi(\vec{r_t},\eta)$.
The boundary condition, $A_\eta (\tau=0,\eta)=0$, cannot be altered by the
gauge transform (\ref{eq:E3.01}). Therefore, we must require that
$\chi(x)=\chi(\vec{r_t})$. Hence, the allowed gauge transform is reduced to
$x$- and $y$-components of the vector potential,
\begin{eqnarray}
A'_r (x)= A_r (x) +\partial_r \chi(\vec{r_t})~.
\label{eq:E3.02} \end{eqnarray} 
The Gauss constraint (\ref{eq:E2.13}),
\begin{eqnarray}
\partial_\eta \partial_{\tau}A_\eta(\tau,\eta)=
-\tau^2 \partial_{r}{\stackrel{\bullet} A}_r(\tau,\eta)
-\tau^2 j^\tau(\tau,\eta)~,\nonumber
\end{eqnarray}
is a hyperbolic differential equation for the function $A_\eta(\tau,\eta)$
which can be integrated (the Riemann function of this equation,
$R(\tau,\eta)=1$). With the boundary conditions, $A_\eta(0,\eta)=0$ and
$A_\eta(\tau,-\infty)=0$, this equation has a unique solution,
\begin{eqnarray}
A_\eta(\tau,\eta)=\int_{-\infty}^{\eta} d\eta~\int_{0}^{\tau} d\tau
[-\tau^2 \partial_{r}{\stackrel{\bullet} A}_r(\tau,\eta)
-\tau^2 j^\tau(\tau,\eta)]~.
\label{eq:E3.03}\end{eqnarray}  
The residual gauge transform (\ref{eq:E3.02}) changes only the integrand
of Eq.~(\ref{eq:E3.03}) 
\begin{eqnarray}
\partial_r A'_r (\tau)~\to~ \partial_r A_r (\tau) +
\Delta_\bot \chi(\vec{r_t})~,\nonumber
\end{eqnarray}
As a consequence of the boundary conditions, the transverse divergence of the
field must vanish at $\tau=0$, $\partial_r A_r (0,\eta)=0$. Therefore,  we must
also have  $$\Delta_\bot \chi(\vec{r_t})=0~,$$ $\chi(x,y)$ must be a harmonic
function. Demanding, that $\chi(x,y)$ vanishes at $|\vec{r_t}|\to \infty$, we
find that $\chi(x,y)=0$. The gauge $A^{\tau}=0$ is fixed completely. The Gauss
law can be unambiguously used to eliminate $A_\eta$ from the list of canonical
variables.

\subsection{Gluon correlators in the gauge $A^{\tau}=0$} 
\label{subsec:SB31}

In this section, we shall write down components of the field correlator 
$\Delta_{10,ij}(x,y)$ in the curvilinear coordinates 
$u=(\tau, \eta,{\vec r})$. We shall denote their covariant components as
$\Delta_{10,ik}(u_1,u_2)$. Later we shall transform them to the standard
Minkowski coordinates and find the correlators of the temporal axial and
the null-plane gauges as their limits in the central rapidity region and
in the vicinity of  the null-planes, respectively. The  most convenient 
(for this purpose) basis consists of the transverse modes  $v^{(\lambda)}$. 
The mode $v^{(1)}$ gives the following contribution to the
correlator $\Delta_{10,ik}$,
\begin{eqnarray}
i \Delta^{(1)}_{10,rs}(1,2)= \int_{-\infty}^{\infty} {d\theta\over 2}
\int {d^2{\vec k}\over (2\pi)^3} 
\bigg(\delta_{rs} -{k_r k_s \over k_{\bot}^{2}}\bigg)   
e^{i{\vec k}({\vec r}_1-{\vec r}_2)}
e^{-ik_{\bot}\tau_1 \cosh (\theta-\eta_1)+ik_{\bot}\tau_2 
\cosh (\theta-\eta_2)}~. 
\label{eq:E3.5} 
\end{eqnarray}           
Realizing that $d\theta/2=dk^3/2k^0$, we recognize a standard representation
of this part of the correlator in terms of the on-mass-shell
plane waves decomposition.

The second part of the correlator is determined by the mode $v^{(2)}$
and has the following components:                                            
\begin{eqnarray}
\Delta^{(2)}_{10,rs}(1,2)= -i\int_{-\infty}^{\infty} {d\theta\over 2}
\int {d^2{\vec k}\over (2\pi)^3} 
{k_r k_s \over k_{\bot}^{2}}   
e^{i{\vec k}({\vec r}_1-{\vec r}_2)} 
f_1(\theta,\tau_1 ,\eta_1)f_{1}^{*}(\theta,\tau_2 ,\eta_2)~,   
\label{eq:E3.6} 
\end{eqnarray}         
\begin{eqnarray}
\Delta^{(2)}_{10,r\eta}(1,2)= -i \int_{-\infty}^{\infty} {d\theta\over 2}
\int {d^2{\vec k}\over (2\pi)^3}   
e^{i{\vec k}({\vec r}_1-{\vec r}_2)}{k_r\over k_{\bot}^2}
 f_1(\theta,\tau_1 ,\eta_1)f_{2}^{*}(\theta,\tau_2 ,\eta_2)
=\Delta^{(2)}_{10,\eta r}(2,1)~, 
\label{eq:E3.7} 
\end{eqnarray}    
\begin{eqnarray}
\Delta^{(2)}_{10,\eta\eta}(1,2)= -i\int_{-\infty}^{\infty}{d\theta\over 2}
\int {d^2{\vec k}\over (2\pi)^3}     
{e^{i{\vec k}({\vec r}_1-{\vec r}_2)}\over k_{\bot}^{2}} 
f_2(\theta,\tau_1 ,\eta_1)f_{2}^{*}(\theta,\tau_2 ,\eta_2)~.   
\label{eq:E3.8} 
\end{eqnarray} 
One may easily see that all components of $\Delta_{10}(1,2)$
vanish when either $\tau_1$ or $\tau_2$ go to zero.         

\subsection{Causal properties of the field commutators 
            in the gauge $A^{\tau}=0$} 
\label{subsec:SB32}

 Causal properties of the radiation field commutator may be studied 
starting from the representation (\ref{eq:E3.3a}). Using Eqs. (\ref{eq:E3.5})
and (\ref{eq:E3.6}) we may conveniently write the contribution of the two
transverse modes in the following form,
\begin{eqnarray}
i \Delta^{(1)}_{0,rs}(1,2)= -i \int {d^2{\vec k}\over (2\pi)^3} 
\bigg(\delta_{rs} - {k_r k_s \over k_{\bot}^{2}}\bigg)  
e^{i{\vec k}{\vec r}}  \int_{-\infty}^{\infty} d\theta\sin k_\bot \Phi_{12}~,
\label{eq:E3.9} 
\end{eqnarray}     
\begin{eqnarray} 
i\Delta^{(2)}_{0,rs}(1,2)= -i \int {d^2{\vec k}\over (2\pi)^3} 
{k_r k_s \over k_{\bot}^{2}} e^{i{\vec k}{\vec r}} 
\int_{-\infty}^{\infty} d\theta \bigg[ 1-
{\cosh 2\eta \over \sinh^2 \theta +\cosh^2 \eta} \bigg]
 (\sin k_\bot \Phi_{12} 
-\sin k_\bot \Phi_1 +\sin k_\bot \Phi_2 )~,  
\label{eq:E3.10} 
\end{eqnarray}        
where we have introduced the following  notation:~
$2\eta=\eta_1-\eta_2~,~~{\vec r}={\vec r}_1 - {\vec r}_2~,~~ 
\Phi_i =\tau_1\cosh (\theta -\eta_i)~,~~ \Phi_{12}=\Phi_1-\Phi_2 ~$.
The sum of (\ref{eq:E3.9}) and (\ref{eq:E3.10}) can be rearranged as
follows,
\begin{eqnarray} 
i\Delta_{0,rs}(1,2)= i\int {d^2{\vec k}~d\theta \over (2\pi)^3} 
e^{i{\vec k}{\vec r}} \bigg[ - \delta_{rs} \sin k_\bot \Phi_{12} +
{k_r k_s \over k_{\bot}^{2}}[ \sin k_\bot \Phi_1- 
\sin k_\bot \Phi_2]\nonumber \\ 
+ k_r k_s \cosh(\eta_1-\eta_2)\int_{0}^{\tau_1}d\tau' 
\int_{0}^{\tau_2}d\tau''~\sin [ k_\bot\tau'\cosh(\theta -\eta)-
k_\bot\tau''\cosh(\theta + \eta)] \bigg]~~. 
\label{eq:E3.10a} 
\end{eqnarray}                                         
Rewriting the integration $d^2{\vec k}~d\theta$ into the three dimensional
integration ~$d^3 {\bf k} / | {\bf k}|$~  in  Cartesian coordinates,
the first integral in (\ref{eq:E3.10a}),  
\begin{eqnarray} 
D_{0}(1,2)= \int {d^2{\vec k}~d\theta \over (2\pi)^3} 
e^{i{\vec k}{\vec r}} \sin k_\bot \Phi_{12} ={{\rm sign}(t_1-t_2) \over 2\pi}
\delta [(t_1-t_2)^2-({\bf r}_1-{\bf r}_2)^2]~, 
\label{eq:E3.11} 
\end{eqnarray}    
is easy to calculate and to recognize  as the commutator of the massless
scalar field. It differs from zero only if the line between the
points $x_1$ and $x_2$ has a light-like direction.
We integrate the first and the third terms in the integrand of
the Eq.~(\ref{eq:E3.10a}) in this way. To reduce the 
two integrals in the second term 
to the same type, we must exclude the factor $1/k_{\bot}^{2}$ using
the fundamental solution of the two-dimensional Laplace operator,
\begin{eqnarray} 
{k_r k_s \over k_{\bot}^{2}} e^{i{\vec k}{\vec r}}=
\partial_r\partial_s \int {d^2{\vec \xi} \over 2\pi} 
\ln |{\vec \xi}-{\vec r}|e^{i{\vec k}{\vec \xi}}~~.
\label{eq:E3.12} 
\end{eqnarray}    
After that, we arrive at the final result,
\begin{eqnarray} 
\Delta_{0,rs}(1,2)= - \delta_{rs} D_{0}(1,2) -
\cosh(\eta_1-\eta_2)\partial_r\partial_s   \int_{0}^{\tau_1}d\tau_1 
\int_{0}^{\tau_2}d\tau_2  D_{0}(1,2) \nonumber \\
+ \partial_r\partial_s \int {d^2{\vec \xi} \over (2\pi)^2}
\ln |{\vec \xi}-{\vec r}| [\delta (\tau_{1}^{2}-{\vec \xi}^{2})-
\delta (\tau_{2}^{2}-{\vec\xi}^{2})]~~. 
\label{eq:E3.13} 
\end{eqnarray}                                         
From this form, it immediately follows that the commutator of the 
potentials vanishes at $\tau_1=\tau_2$.  An even stronger result
is found for the commutator of the two electric fields,
\begin{eqnarray}
 [ E_r(1), E_s(2)] = 
{\partial^2 \over \partial \tau_1 \partial \tau_2 }
i\Delta_{0,rs}(1,2)= 
\bigg[ - \delta_{rs} {\partial^2 \over \partial \tau_1 \partial \tau_2 }
- \cosh(\eta_1-\eta_2) 
{\partial^2 \over \partial x^r \partial x^s } \bigg]~iD_{0}(1,2)~~.
\label{eq:E3.14}\end{eqnarray}                                
This commutator vanishes everywhere except on the light cone, in full
compliance with the microcausality principle for the electric field
which is an observable. However, this does not happen for the commutator
of the potentials since they are defined non-locally.  It 
vanishes neither at space-like  nor at time-like separation
because the line of integration which recovers the potential at the point
$x_2$, in general, intersects ({\em e.g.} at some point $x_3$) with the light 
cone which has its vertex at the point $x_1$, and the commutator of the
electric fields at the points $x_1$ and $x_3$ is not zero.

Similar results take place for the commutator of the $\eta$-components of
the potential and the electric field. The field commutator,
\begin{eqnarray}
 [ E_\eta (1), E_\eta(2)] = 
{\partial^2 \over \partial \tau_1 \partial \tau_2 }
i\Delta_{0,\eta\eta}(1,2)= -i \nabla_{\bot}^{2} D_{0}(1,2)~~,
\label{eq:E3.15}\end{eqnarray}                                
is entirely causal, while the commutator of the potentials,
\begin{eqnarray}
 [ A_\eta (1), A_\eta (2)]=i\Delta_{0,\eta\eta}(1,2)=-i \nabla_{\bot}^{2} 
\int_{0}^{\tau_1}\tau_1d\tau_1\int_{0}^{\tau_2}\tau_2d\tau_2D_{0}(1,2)~~,
\label{eq:E3.16}\end{eqnarray}    
does not vanishes at space-like distances, except for $\tau_1=\tau_2$.
Finally, the formally designed commutator between the $r$-- and 
$\eta$--components of the electric field  (the two observables),
\begin{eqnarray}
 [ E_r (1), E_\eta(2)] = 
{\partial^2 \over \partial \tau_1 \partial \tau_2 } 
i\Delta_{0,r\eta}(1,2)= - {\partial^2 \over \partial x^r \partial \eta } 
\bigg( {\tau_2\over\tau_1}D_{0}(1,2)\bigg)~~,
\label{eq:E3.17}\end{eqnarray}      
is entirely confined to the light cone, while the commutator of the
potentials (which are not the observables),
\begin{eqnarray}
 [ A_r (1), A_\eta(2)] = i\Delta_{0,r\eta}(1,2)=
-\int_{0}^{\tau_1} {d \tau_1\over\tau_1} \int_{0}^{\tau_2}\tau_2 d \tau_2  
 {\partial^2 \over \partial x^r \partial \eta } D_{0}(1,2)~~,
\label{eq:E3.18}\end{eqnarray}      
does not vanish at the space-like distance, even at $\tau_1=\tau_2$.
This result, however, is not a subject for any concern since the potentials
are defined non-locally and commutation relations for 
the electric and magnetic
( cf. (\ref{eq:E2.15A}) ) fields are reproduced correctly. Moreover, we have
argued above that the $\eta$-components of $A$ and $E$ are not the
canonical variables since the constraints express them via the $x$- and
$y$-components.

The ``acausal'' behavior of the Riemann function, 
~$\Delta_{0}^{\mu\nu}(1,2)$,   may cause doubts 
whether the gauge $A^\tau =0$ allows for meaningful retarded
and advanced Green functions which, by causality, should vanish at
space-like distances. Fortunately, this anomalous behavior appears only
for the gauge--variant potential; the response functions for observable
electric and magnetic fields are causal. This can be easily seen,  {\em
e.g.}, from Eqs. (\ref{eq:E2.19}), (\ref{eq:E2.23})  and (\ref{eq:E2.24}),
which are the usual inhomogeneous relativistic wave 
equations for the various physical
components of the field strengths ${\cal E}$ and  ${\cal B}$.

\subsection{Canonical commutation relations in the gauge $A^{\tau}=0$} 
  \label{subsec:SB33}

A proof of the commutation relations (\ref{eq:E3.2}) for the Fock 
operators follows the
standard guidelines \cite{Bjorken}. First, the creation and annihilation
operators are {\em defined} via the relations,
\begin{eqnarray}
c_{\lambda}(\nu,{\vec k})= (V^{(\lambda)}_{\nu {\vec k}},A)=
i{\bf g}^{ij}\int d^3{\bf x}
[ V^{(\lambda)\ast }_{\nu {\vec k};j}(x){\stackrel{\bullet} A}_i({\bf x},\tau)
-{\stackrel{\bullet} V}^{(\lambda)\ast }_{\nu {\vec k};j}(x)
                            A_i({\bf x},\tau)]~, \nonumber \\
c^{\dag}_{\lambda}(\nu,{\vec k}) =(A,V^{(\lambda)}_{\nu {\vec k}})=
i{\bf g}^{ij}\int d^3{\bf x}
[A_i({\bf x},\tau){\stackrel{\bullet} V}^{(\lambda) }_{\nu {\vec k};j}(x)
-{\stackrel{\bullet} A}_i({\bf x},\tau)V^{(\lambda) }_{\nu {\vec k};j}(x)]~~. 
\label{eq:E3.19}\end{eqnarray}  
This results in the following expression for the commutator,
\begin{eqnarray}
[c_{\lambda}(\nu,{\vec k}),c^{\dag}_{\lambda'}(\nu',{\vec k'})] =
\int d^3{\bf x}d^3{\bf y} {\rm g}^{ij}(x){\rm g}^{lm}(y)~
\big\{ [A_i({\bf x},\tau),{\stackrel{\bullet} A}_l({\bf y},\tau)]~
\big({\stackrel{\bullet} V}^{(\lambda)\ast }_{\nu {\vec k};j}(x)
V^{(\lambda')}_{\nu'{\vec k}';n}(y)
-{\stackrel{\bullet} V}^{(\lambda')\ast}_{\nu' {\vec k}';j}(x)  
V^{(\lambda)}_{\nu{\vec k};n}(y)\big) \nonumber \\ 
+[A_i({\bf x},\tau),A_l({\bf y},\tau)]~
{\stackrel{\bullet} V}^{(\lambda)\ast }_{\nu {\vec k};j}(x)
{\stackrel{\bullet} V}^{(\lambda')}_{\nu'{\vec k}';n}(y) +
[{\stackrel{\bullet} A}_i({\bf x},\tau),{\stackrel{\bullet} A}_l({\bf y},\tau)]~
 V^{(\lambda')\ast}_{\nu' {\vec k}';j}(x)
V^{(\lambda)}_{\nu{\vec k};n}(y) \big\} ~~. 
\label{eq:E3.20}\end{eqnarray} 
Most of the terms in the second line vanish due to the commutation
relations. Next, we rely on the following guess about the form of the 
commutator,    
\begin{eqnarray}
[A_i(x), A_j(y)]=\sum_{\lambda=1,2}\int d\nu ~d^2{\vec k}~
\big( V^{(\lambda)}_{\nu{\vec k};i}(x)V^{(\lambda)\ast }_{\nu{\vec k};j}(y)-
V^{(\lambda)\ast}_{\nu{\vec k};i}(x)
V^{(\lambda)}_{\nu {\vec k};i}(y)\big) ~~,
\label{eq:E3.22}\end{eqnarray} 
which leads to the proper equal-time commutation relations for the
independent canonical variables.
Finally, explicitly using the orthogonality relations for the eigen-modes
$V^{(\lambda)}$, we immediately obtain the commutation relations 
(\ref{eq:E3.2}).  

\section{Longitudinal propagator and static fields} 
\label{sec:SN5}
         
In this section, we shall find the explicit expressions for the kernels
(\ref{eq:E2.40}) and (\ref{eq:E2.44}) which represent the longitudinal and 
instantaneous components of the gauge field produced by the ``external''
current $j^\mu$. The calculations are lengthy and their details can be
found in Appendix~3. Here, we present only the final answers. 

The components of the longitudinal propagator
$\Delta^{(L)}_{lm}(\tau_2,{\vec r},\eta)$ are already obtained in the form 
of the three-dimensional integrals (\ref{eq:E2.40}). $\Delta^{(L)}$
depends on the differences of the curvilinear spatial coordinates,
${\vec r}={\vec r}_1-{\vec r}_2$ and $\eta=\eta_1-\eta_2$, but {\em not}
on the difference of the temporal arguments $\tau_1$ (of the field) and
$\tau_2$ (of the source).  Introducing the shorthand notation for the 
distance in the $(xy)$--plane, $r_{\bot}=|{\vec r}|$, and for the full
distance $R_2=R(\tau_2)=[({\vec r}_1-{\vec r}_2)^2+
\tau_{2}^{2}\sinh^2(\eta_1-\eta_2)]^{1/2}$ 
between the two points of the surface $\tau_2=const$, we obtain:
\begin{eqnarray} 
\Delta^{(L)}_{rs}= -{\theta(\tau_2-r_{\bot}) \over 4\pi}
\bigg[{1\over r_{\bot}^{2}} \bigg(1-{\tau_2\cosh\eta\over R_2}\bigg)
\bigg(\delta_{rs}- {2 x^r x^s\over r_{\bot}^{2}}\bigg)  
-{2 x^r x^s\over r_{\bot}^{2}} 
{\tau_2\cosh\eta\over R_{2}^{3}}\bigg]~,\nonumber\\
\Delta^{(L)}_{\eta s}= -{\theta(\tau_2-r_{\bot}) \over 4\pi}
{x^s\over r_{\bot}^{2}}{\tau_2\sinh\eta\over R_2}
{\tau_{2}^{2}-r_{\bot}^{2}\over R_{2}^{2}}~,~~~ 
\Delta^{(L)}_{r \eta}= -{\theta(\tau_2-r_{\bot}) \over 4\pi}
{x^r \over r_{\bot}^{2}}{\tau_{2}^{3}\sinh\eta\over R_{2}^{3}}~,  \nonumber\\
\Delta^{(L)}_{\eta\eta}= {\tau_{2}^{2}\over 2}\delta({\vec r})\delta(\eta)
+{\theta(\tau_2-r_{\bot}) \over 4\pi}
\bigg[ 2{\eta\coth\eta-1 \over\sinh^2\eta}  +
{\tau_2  \cosh\eta \over R_2 \sinh^2\eta}
\bigg( 3 -{r_{\bot}^{2}\over R_{2}^{2}}\bigg)  -
{2 \cosh\eta \over \sinh^3|\eta|}~L_2 \bigg]~~.
\label{eq:E5.1}\end{eqnarray}                               
where $~L_2=L(\tau_2)=\ln [(\tau_2\sinh |\eta|+R_2)/r_{\bot}]$.  
By  examination of Eq.~(\ref{eq:E2.39}), one may see that after the 
replacement of $\tau_2$ by $\tau_1$,  the same kernel, 
$\Delta^{(L)}_{lm}(\tau_1,{\vec r}, \eta)$, determines the components
$ E_{m}^{(L)}(\tau_1)$ of the longitudinal part of the electric field via 
the components $j^m(\tau_1)$ of the current at the same time.
These propagators do not respect the light cone, but have a remarkable 
property that the longitudinal fields at the surface of the constant
proper time $\tau$ do not exist at the distance $r_\bot$ from their sources  
that exceed $\tau$. This establishes the upper limit for the possible
dynamical correlations between the longitudinal fields in the $(xy)$-plane.     

One more representation of the longitudinal propagator is interesting at least
in two respects. First, for the practical calculations in the paper [IV], we
shall need the longitudinal part of the propagator in the mixed representation,
$\Delta^{(long)}_{lm} (\tau_1,\tau_2;\eta_1-\eta_2,{\vec k}_t)$, which can be
shown to be
\begin{eqnarray}
 \Delta^{[long]}_{rs}= {k_rk_s\over k_t^2}
 \bigg\{ - {k_t\cosh |\eta|\over 2} 
 \int_{\tau_2}^{\tau_1}  e^{-t k_t\sinh |\eta|} dt 
 - \int {d\alpha\over 2\pi}
 \tanh(\alpha+{\eta\over 2}) \tanh(\alpha-{\eta\over 2})
\big[~\sin k_tT_1-\sin k_tT_2~\big]~\bigg\},
\label{eq:E5.2}
\end{eqnarray} 
\begin{eqnarray}
 \Delta^{[long]}_{\eta\eta}= -  {\tau_1^2-\tau_2^2\over 2}\delta(\eta)
 + {k_t\cosh |\eta|\over 2} 
 \int_{\tau_2}^{\tau_1}  e^{-t k_t\sinh |\eta|} t^2 dt 
 \hspace{4cm}  \nonumber\\
 -\int {d\alpha\over 2\pi}
 { \sin k_tT_1-\sin k_tT_2-k_tT_1\cos k_tT_1 
 +k_tT_2\cos k_tT_2 \over k_t^2\cosh^2(\alpha+{\eta\over 2})
 \cosh^2(\alpha-{\eta\over 2})} ,
\label{eq:E5.3}
\end{eqnarray} 
where the last integral terms in Eqs.~(\ref{eq:E5.2}) and 
(\ref{eq:E5.3}) provide that
the longitudinal part of the field obey the boundary condition we imposed  at
$\tau\to 0$. These terms cancel with the similar terms in the radiation part of
the retarded propagator $\Delta_{[ret]}(\tau_1,\tau_2)$. Eventually, the last
fact guarantee that acausal terms (which were the subject for concern in the 
course of  the canonical quantization  in Sec.~\ref{subsec:SB33}) do not
contribute to the  dispersion equation that we derive in the next paper.
Second, the first very simple by its structure contact term in the component 
$D^{[long]}_{\eta\eta}$ of the propagator, 
\begin{eqnarray} 
\Delta^{[contact]}_{\eta\eta}(\tau_2,\tau_1;\eta_2-\eta_1;\vec{k_t})
=-{\tau_1^2-\tau_2^2 \over 2}\delta (\eta)~,
\label{eq:E5.4}\end{eqnarray}
appears to be the only part of the enormously complicated  full retarded
propagator $\Delta^{[ret]}_{lm}$ which significantly contributes the amplitude
of the forward quark-quark scattering at the earliest stage of the collision.

\vspace{1cm}

\noindent {\bf ACKNOWLEDGMENTS}

The author is grateful to Berndt Muller,Edward Shuryak and Eugene Surdutovich
for helpful discussions at various stages in the development of this work,
and appreciate the help of Scott Payson who critically read the
manuscript.

This work was partially supported by the U.S. Department of Energy under
Contract  No. DE--FG02--94ER40831.

\bigskip

\bigskip   
\bigskip   
\appendix
\renewcommand{\theequation}{A\arabic{equation}}
\setcounter{equation}{0}  

\centerline {\bf APPENDIX ~A. Modes of the free gauge field}    

\bigskip

Here, we shall obtain the complete set of the one-particle solutions  to
the homogeneous system  of the Maxwell equations with the gauge 
$A^{\tau}=0$, that is, Eqs.~(\ref{eq:E2.10})
and (\ref{eq:E2.11}). This gauge 
condition explicitly depends on the coordinates, thus introducing
effective non-locality in the path integral that represents the action.
Therefore, it becomes impossible to invert the differential operators using
the standard symbolic methods.  An explicit form of the one-particle
solutions  becomes necessary in order to find  the Wightman functions of
the free vector field, to establish the the explicit form of the field
commutators, and to separate  the propagators of the transverse and the
longitudinal fields.
It is natural to look for the solution in the form of the Fourier
transform with respect to the spatial coordinates,
\begin{eqnarray}  
 A_i (x) = \int_{-\infty}^{\infty} d \nu\int d^2 {\vec k} ~ e^{i\nu\eta}
e^{i{\vec k}{\vec r}} ~A_{i} ({\vec k},\nu ,\tau)~.
 \label{eq:A1}\end{eqnarray} 
Then, the system of  second order ordinary 
differential  equations for the Fourier transforms takes the following
form:
\begin{eqnarray}
[\tau\partial_{\tau}^{2}+\partial_{\tau}+
{\nu^2\over \tau} +\tau k_{y}^{2}]A_x({\vec k},\nu ,\tau)
-\tau k_{x} k_{y}A_y ({\vec k},\nu ,\tau)- 
{\nu k_x\over \tau}A_\eta ({\vec k},\nu ,\tau) =0~~,  
\label{eq:A2}\end{eqnarray}         
\begin{eqnarray}
-\tau k_{x} k_{y}A_x ({\vec k},\nu ,\tau)+
[\tau\partial_{\tau}^{2}+\partial_{\tau}+
{\nu^2\over \tau} +\tau k_{x}^{2}]A_y({\vec k},\nu ,\tau)-
{\nu k_y\over \tau}A_\eta ({\vec k},\nu ,\tau) =0~~,
\label{eq:A3}\end{eqnarray}               
\begin{eqnarray}
-{\nu k_x\over \tau}A_x ({\vec k},\nu ,\tau)-
{\nu k_y\over \tau}A_y ({\vec k},\nu ,\tau) +
[{1\over \tau}\partial_{\tau}^{2}-{1\over \tau^2}\partial_{\tau}+
{1\over \tau}k_{\bot}^{2}]A_\eta ({\vec k},\nu ,\tau) =0~.
\label{eq:A4}\end{eqnarray}         
In this form, the system is manifestly symmetric and self-adjoint.
An additional equation of the constraint reads as
\begin{eqnarray}
{\cal C}({\vec k},\nu ,\tau)=
 {1\over \tau}\nu\partial_{\tau}A_\eta+
\tau\partial_{\tau} [k_x A_x({\vec k},\nu ,\tau) 
+k_{y} A_y({\vec k},\nu ,\tau)] = 0~~.
\label{eq:A5}\end{eqnarray}          
Let us rewrite the homogeneous system of the Maxwell equations
in terms of the variables 
\begin{eqnarray}   
\Phi= \partial_{x}A_x +\partial_{y} A_y ,\;\;\; 
\Psi= \partial_{y}A_x -\partial_{x} A_y ,\;\;\;  {\rm and} \;\; 
{\sf A}=A_\eta .
\label{eq:A6}\end{eqnarray}    
One immediately sees that the equation for the Fourier component
$\Psi({\vec k},\nu ,\tau)$ of the longitudinal magnetic field
$\Psi(x)$ decouples,
\begin{eqnarray}
[\partial_{\tau}^{2}+{1\over \tau}\partial_{\tau}+
{\nu^2 \over \tau^2} +k_{\bot}^{2}]\Psi({\vec k},\nu ,\tau)=0~.
\label{eq:A7}\end{eqnarray}       
Then, the other two equations of motion  take shape, {\em i.e},
\begin{eqnarray}  
[\tau^2 \partial_{\tau}^{2}+\tau\partial_{\tau}+
\nu^2 ]\Phi_{{\vec k},\nu}(\tau)-i\nu k_{\bot}^2 
{\sf A}_{{\vec k},\nu}(\tau) =0~,
\label{eq:A8}\end{eqnarray}        
\begin{eqnarray}  
[\partial_{\tau}^{2}-{1\over\tau}\partial_{\tau}+
k_{\bot}^{2}] {\sf A}_{{\vec k},\nu}(\tau) 
+i\nu \Phi_{{\vec k},\nu}(\tau) =0~. 
\label{eq:A9}\end{eqnarray}  
The additional constraint equation can be conveniently rewritten as     
\begin{eqnarray}
{\cal C}({\vec k},\nu ,\tau)=
 {i\nu\over \tau} \partial_\tau {\sf A}_{{\vec k},\nu}(\tau) 
+\tau\partial_\tau \Phi_{{\vec k},\nu}(\tau) = 0~~.
\label{eq:A10}\end{eqnarray} 
This is an independent equation. However, the conservation of the
constraint along the Hamiltonian time $\tau$ is a consequence of
the equations of motion, and it {\it can} be employed  to obtain
the independent equations for the components of the vector field.        
This is easily done in terms of the auxiliary functions,
\begin{eqnarray}
\varphi_{{\vec k},\nu}(\tau)=\tau{\stackrel{\bullet} \Phi}_{{\vec k},\nu}(\tau)\;\;\; 
{\rm and}\;\;\;a_{{\vec k},\nu}(\tau)=
\tau^{-1}{\stackrel{\bullet} {\sf A}}_{{\vec k},\nu}(\tau),
\label{eq:A11}\end{eqnarray}  
which are directly connected to the ``physical'' components of the
electric field, ${\cal E}^m = \sqrt{-{\rm g}}{\rm g}^{ml}{\stackrel{\bullet} A}_l$~:
\begin{eqnarray}
\partial_\tau [\partial_{\tau}^{2}+{1\over \tau}\partial_{\tau}+
{\nu^2 \over \tau^2} +k_{\bot}^{2}]\varphi_{{\vec k},\nu}(\tau)=0~~,
\label{eq:A12}\end{eqnarray}                 
\begin{eqnarray}
\partial_\tau [\tau^2 \partial_{\tau}^{2}+\tau\partial_{\tau}+
\nu^2  +k_{\bot}^{2}\tau^2 ] a_{{\vec k},\nu}(\tau)=0~~,
\label{eq:A13}\end{eqnarray} 
As a result, we see that  the functions $\varphi_{{\vec k},\nu}(\tau)$
and $a_{{\vec k},\nu}(\tau)$ obey inhomogeneous Bessel equations,
\begin{eqnarray}
[\partial_{\tau}^{2}+{1\over \tau}\partial_{\tau}+
{\nu^2 \over \tau^2} + k_{\bot}^{2}]\varphi_{{\vec k},\nu}(\tau)
={\vec k}^{2}c_\varphi ~~,
\label{eq:A14}\end{eqnarray}             
\begin{eqnarray}
[\partial_{\tau}^{2}+{1\over \tau}\partial_{\tau}+
{\nu^2 \over \tau^2} +k_{\bot}^{2}] a_{{\vec k},\nu}(\tau)
=\tau^{-2} c_a ~~,              
\label{eq:A15}\end{eqnarray}             
where $c_\varphi$ and $c_a$  are arbitrary constants.
We may now cast the solution of these equations in the
form of the sum of the partial solution of the inhomogeneous equation
and a general solution of the homogeneous equation,
\begin{eqnarray}
\Psi_{{\vec k},\nu}(\tau) = a  H^{(2)}_{-i\nu}(k_{\bot}\tau) +
a^*  H^{(1)}_{-i\nu}(k_{\bot}\tau)~~,
\label{eq:A16a}\end{eqnarray}       
\begin{eqnarray}
\varphi_{{\vec k},\nu}(\tau) = c  H^{(2)}_{-i\nu}(k_{\bot}\tau) +
c^*  H^{(1)}_{-i\nu}(k_{\bot}\tau) + c_\varphi s_{1,i\nu}(k_{\bot}\tau)~,
\label{eq:A16}\end{eqnarray}             
\begin{eqnarray}
a_{{\vec k},\nu}(\tau) = \gamma  H^{(2)}_{-i\nu}(k_{\bot}\tau) +
\gamma^*  H^{(1)}_{-i\nu}(k_{\bot}\tau) + c_a s_{-1,i\nu}(k_{\bot}\tau)~,
\label{eq:A17}\end{eqnarray}             
where $s_{\mu,\nu}(x)$ is the so-called Lommel function \cite{TF,Wat}. 

Furthermore, it is useful to notice that  the system of the Maxwell equations 
(\ref{eq:E2.10}) and (\ref{eq:E2.11}) also has an infinite set of the
$\tau$-independent solutions of the form
\begin{eqnarray}
 W_i(\eta,{\vec r}) =\partial_i \chi (\eta,{\vec r})~~,
\label{eq:A19}\end{eqnarray}    
where $\chi$ is an arbitrary function of the spatial coordinates 
$\eta$ and ${\vec r}$.
Thus, they are the pure gauge solutions of the Abelian theory, compatible
with the gauge condition.     

In order to find the coefficients one should integrate Eqs. (\ref{eq:A16})
and (\ref{eq:A17}) with respect to the Hamiltonian time $\tau$, thus
finding the functions $\Phi$ and ${\sf A}$. Next, it is necessary to solve 
Eqs.(\ref{eq:A6}) for the Fourier components of the vector potential and to
substitute them into the original system of 
Eqs.~(\ref{eq:A2})--(\ref{eq:A5}).  Using functional relations from 
Appendix 2, one obtains that ~$c+\nu\gamma=0$ ~and ~$c_a-\nu c_\varphi=0$~.

One of the solutions, (already normalized according to Eq.(\ref{eq:E2.14}))
is found immediately: 
\begin{eqnarray}  
V^{(1)}_{{\vec k},\nu}(x)={e^{-\pi\nu/2}\over 2^{5/2}\pi k_{\bot}} 
\left( \begin{array}{c} 
                         k_y \\ 
                        -k_x \\ 
                         0 
                             \end{array} \right)
H^{(2)}_{-i\nu} (k_{\bot}\tau) e^{i\nu\eta +i{\vec k}{\vec r}}~~.
\label{eq:A18} 
\end{eqnarray} 
Initially, the components of the vector mode $V^{(2)}$
appear in the following form required by the convergence of the integral,
\begin{eqnarray}
 \left( \begin{array}{c}
            \nu  k_r R^{(2)}_{-1,-i\nu}(k_{\bot}\tau |S) \\
          -R^{(2)}_{1,-i\nu}(k_{\bot}\tau | s) -
                 i \nu [e^{\pi\nu /2}/ \sinh(\pi\nu /2)]
                               \end{array} \right)
e^{i\nu\eta +i{\vec k}{\vec r}}~. \nonumber
\end{eqnarray}                                     
However, it can be gauge transformed to the more compact form,
\begin{eqnarray}  
V^{(2)}_{{\vec k},\nu}(x)={e^{-\pi\nu/2}\over 2^{5/2}\pi k_{\bot}} 
\left( \begin{array}{c} 
       k_r \nu R^{(2)}_{-1,-i\nu}(k_{\bot}\tau |s)  \\ 
       - R^{(2)}_{1,-i\nu}(k_{\bot}\tau | s) 
                                            \end{array} \right)
                e^{i\nu\eta +i{\vec k}{\vec r}}~. 
\label{eq:A25}\end{eqnarray} 
The third solution  (the last one by the count of the non-vanishing
components of the vector potential in the gauge $A^\tau =0$)
has the following form,   
\begin{eqnarray} 
V^{(3)}_{{\vec k},\nu}(x)=
    \left( \begin{array}{c} 
                 k_r  Q_{-1,i\nu}(k_{\bot}\tau) \\ 
                 \nu Q_{1,i\nu}(k_{\bot}\tau)
                                            \end{array} \right)
                   e^{i\nu\eta +i{\vec k}{\vec r}}~~.    
\label{eq:A26}\end{eqnarray}         

The modes $V^{(1)}$ and $V^{(2)}$ are the normalized solutions of the 
Maxwell equations. They are orthogonal and obey the normalization condition,
\begin{eqnarray} 
(V^{(1,2)}_{{\vec k},\nu},V^{(1,2)}_{{\vec k}',\nu'})=\delta(\nu-\nu')
\delta({\vec k}-{\vec k}'),\;\;\;
(V^{(1)}_{{\vec k},\nu},V^{(2)}_{{\vec k},\nu})=0~~.
\label{eq:A28}\end{eqnarray}  
which can be easily verified by means of Eq.~(\ref{eq:A2.5}). 
The norm of these solutions is given by the Eq.~(\ref{eq:E2.14}).         
A normalization coefficient of the mode $V^{(3)}$ 
is not defined, as this mode has a zero norm. It is also
orthogonal to $V^{(1)}$ and  $V^{(2)}$:
\begin{eqnarray} 
(V^{(3)}_{{\vec k},\nu},V^{(3)}_{{\vec k}',\nu'})=
(V^{(1)}_{{\vec k},\nu},V^{(3)}_{{\vec k},\nu})=
(V^{(2)}_{{\vec k},\nu},V^{(3)}_{{\vec k},\nu})=0~~.
\label{eq:A30}\end{eqnarray}   
Thus this mode drops out from the decomposition of the free gauge field. 
 
The conservation of the constraint can be obtained as a consequence of the 
Eqs. ~(\ref{eq:A12}) and (\ref{eq:A13}) in the form,
\begin{eqnarray}
\tau \partial_\tau [\varphi_{{\vec k},\nu}(\tau)+
\nu a_{{\vec k},\nu}(\tau)] 
\equiv \tau\partial_\tau {\cal C}_{{\vec k},\nu}(\tau)=0~~,
\label{eq:A31}\end{eqnarray} 
which reassures us of the consistency between the dynamic equations and
conservation of the Gauss' law constraint.
  
One can explicitly check that the modes $V^{(1)}$ and $V^{(2)}$  obey
the constraint equation (\ref{eq:A10}), which expresses  Gauss' law.
The mode $V^{(3)}$ does not. This mode corresponds to the longitudinal
field which cannot exist without the source.

\renewcommand{\theequation}{B\arabic{equation}}
\setcounter{equation}{0}  

\bigskip

\centerline {\bf APPENDIX ~B. Mathematical miscellany}       

\bigskip

This appendix contains a list of mathematical formulae for the functions
which appear in various calculations in the body of the paper and Appendix~1.
The components of the vector field are expressed via  two types of
integrals. The first of them was studied in Ref.\cite{Wat}:
\begin{eqnarray}   
R^{(j)}_{\mu,\nu}(x |{\sf S})=\int x^\mu H^{(j)}_{\nu}(x) d x =
x[(\mu +\nu -1)H^{(j)}_{\nu}(x) {\sf S}_{\mu -1,\nu -1}(x)-
H^{(j)}_{\nu -1}(x){\sf S}_{\mu ,\nu}(x)]~~,
\label{eq:A2.1}\end{eqnarray}    
where ${\sf S}_{\mu ,\nu}$ stands for any of the two Lommel functions, 
$s_{\mu ,\nu}$ or $S_{\mu ,\nu}$ ~\cite{TF,Wat}. [ Whenever we omit the 
indicator $|{\sf S})$, the function $R^{(j)}_{\mu ,\nu}(x |s)$ is 
assumed.]  The second type of  integrals,  
\begin{eqnarray}   
Q_{\mu,\nu}(x) = \int_{0}^{x} x^\mu  d x s_{-\mu,\nu}(x)~, 
\label{eq:A2.2}\end{eqnarray} 
is a new one.
The functions $R^{(j)}_{\mu,\nu}(x|{\sf S})$  are introduced 
as  indefinite integrals. The preliminary choice of the lower limit
and, consequently, the choice of which of the functions,
$s_{\mu ,\nu}$ or $S_{\mu ,\nu}$,is used
is motivated by the requirement of convergence and regular behavior.
One can easily prove that 
\begin{eqnarray}   
R^{{2 \choose 1}}_{-1,\mp i\nu}(x |S)-R^{{2 \choose 1}}_{-1,\mp i\nu}(x |s) = 
 {\mp i e^{\pi\nu /2} \over \nu \sinh(\pi\nu /2)}, \nonumber \\
R^{{2 \choose 1}}_{1,\mp i\nu}(x |S)-R^{{2 \choose 1}}_{1,\mp i\nu}(x |s) = 
 {\pm i \nu e^{\pi\nu /2} \over \sinh(\pi\nu /2)}.
\label{eq:A2.3}\end{eqnarray}  
We often use the following relation between  Lommel functions 
\cite{TF,Wat},
\begin{eqnarray}
{\sf S}_{1,i\nu}(k_{\bot}\tau)=1- \nu^2 {\sf S}_{-1,i\nu}(k_{\bot}\tau)~~.     
\label{eq:A2.4}\end{eqnarray} 
From the integral representations (\ref{eq:A2.1}) and  (\ref{eq:A2.2}),
it is straightforward to derive the functional relations
\begin{eqnarray}
R^{(j)}_{-1,i\nu}(k_{\bot} \tau)+ 
{1 \over \nu^2}R^{(j)}_{1,i\nu}(k_{\bot} \tau) 
= - {\tau \over \nu^2}{\partial \over \partial \tau}
H^{(j)}_{i\nu}(k_{\bot}\tau)~~,
\label{eq:A2.5}\end{eqnarray}
\begin{eqnarray}
Q_{-1,i\nu}(k_{\bot} \tau)-Q_{1,i\nu}(k_{\bot} \tau)=
-{\tau \over \nu^2} {\partial\over\partial\tau} s_{1,i\nu}(k_{\bot}\tau)=
\tau {\partial\over\partial\tau} s_{-1,i\nu}(k_{\bot}\tau)~~.
\label{eq:A2.6}\end{eqnarray} 
The Wronskian of the Hankel and Lommel functions,
\begin{eqnarray}   
W\{s_{1 ,i\nu}(x), H^{(j)}_{i\nu}(x)\} =
 - {1 \over x}  R^{(j)}_{1,i\nu}(x)~~,
\label{eq:A2.7}\end{eqnarray}       
must be obtained in order
to prove orthogonality of $V^{(2)}$ and $V^{(3)}$. To prove
(\ref{eq:A2.7}), one should use the following representation for the Lommel
function,
\begin{eqnarray}   
s_{1 ,i\nu}(x)= {\pi\over 4i}[ H^{(1)}_{i\nu}(x)R^{(2)}_{1,i\nu}(x)
- H^{(2)}_{i\nu}(x)R^{(1)}_{1,i\nu}(x)]~~,
\label{eq:A2.8}\end{eqnarray}       
which follows from Eq.~(\ref{eq:A2.1}) and, consequently,
\begin{eqnarray}   
s'_{1 ,i\nu}(x)= {\pi\over 4i}[ H^{(1)'}_{i\nu}(x)R^{(2)}_{1,i\nu}(x)
- H^{(2)'}_{i\nu}(x)R^{(1)}_{1,i\nu}(x)]~~.
\label{eq:A2.9}\end{eqnarray}   
In order to prove relation (\ref{eq:E2.30}) one should use the representation
(\ref{eq:A2.1}) for the functions $R^{(j)}_{\mu,\nu}$ and the Wronskian
of the two independent Hankel functions. 
The proof of relation (\ref{eq:E2.41})
begins with replacing the functions $R^{(j)}_{-1,i\nu}$ by  
$R^{(j)}_{1,i\nu}$ by means of Eq.~(\ref{eq:A2.5}). The final result 
follows from Eq.~(\ref{eq:A2.9}) and (\ref{eq:A2.6}).

\bigskip

\renewcommand{\theequation}{C\arabic{equation}}
\setcounter{equation}{0}  

\centerline {\bf APPENDIX ~C. Calculation of the longitudinal part
            of the propagator} 

\bigskip
      
The kernels (\ref{eq:E2.42}) and (\ref{eq:E2.44}) of the longitudinal
and instantaneous parts of the propagator are given in the form of the
three--dimensional Fourier integrals $d\nu d^2{\vec k}$. Here, we describe
the major steps of the calculations which lead to Eqs.~(\ref{eq:E5.1}).

We permanently use the following integral representation for the Hankel 
functions,
\begin{eqnarray}
e^{-\pi\nu/2}e^{\pm i\nu\eta} H^{{2 \choose 1}}_{\mp i\nu}
(k_{\bot}\tau)= {\pm i\over \pi} \int_{-\infty}^{\infty} 
e^{\mp ik_{\bot}\tau \cosh (\theta-\eta)} e^{\pm i\nu\theta} d \theta~~,
\label{eq:A3.1}
\end{eqnarray}         
which allows one to calculate many integrals by changing the order
of integration.
The Lommel function $S_{1,i\nu}$ has a similar representation,
\begin{eqnarray} 
S_{1,i\nu}(x)= x\int_{0}^{\infty} \cosh u \cos\nu u ~e^{-x\sinh u} d u~.
\label{eq:A3.2}\end{eqnarray}  
Integrating by parts, and using Eq.~(\ref{eq:A2.4}), we find the
integral representation for $S_{-1,i\nu}$, 
\begin{eqnarray} 
\nu S_{-1,i\nu}(x)= \int_{0}^{\infty}  \sin (\nu u) ~e^{-x\sinh u} d u~.
\label{eq:A3.3}\end{eqnarray}  
We start with the integral representation (\ref{eq:A2.2}) of the functions 
$Q_{\pm 1,i\nu}$ and perform an integration over $\nu$.     
To compute the integrals from the function $s_{1,i\nu}$ it can be
conveniently decomposed in the following way,
\begin{eqnarray} 
s_{1,i\nu}(x)= S_{1,i\nu}(x)-h_{i\nu}(x)~~,~~~ 
h_{i\nu}(x)={e^{-\pi\nu /2}\over 2}{\pi\nu /2 \over\sinh(\pi\nu /2 )} 
[H^{(1)}_{i\nu}(x)+ H^{(2)}_{-i\nu}(x)]~,
\label{eq:A3.4}\end{eqnarray}        
which allows one to find 
\begin{eqnarray} 
\int_{-\infty}^{\infty}S_{1,i\nu}(k_{\bot}\tau)e^{i\nu\eta}d\nu= 
\pi k_{\bot}\tau\cosh\eta e^{-k_{\bot}\tau\sinh |\eta|}~~,\nonumber\\ 
\int_{-\infty}^{\infty} \nu S_{-1,i\nu}(k_{\bot}\tau) e^{i\nu\eta}d\nu= 
i\pi {\rm sign}\eta e^{-k_{\bot}\tau\sinh |\eta|}~~. 
\label{eq:A3.5}\end{eqnarray}  
The similar Fourier integrals from the function $h_{i\nu}$ are calculated
using the representation (\ref{eq:A3.1}) for the Hankel functions 
and the integral,
\begin{eqnarray} 
{\pi\nu/2 \over \sinh(\pi\nu/2 ) }= {1\over 2}
\int_{-\infty}^{\infty} d\theta{e^{i\nu\theta}\over \cosh^2\theta}~~.
\label{eq:A3.6}\end{eqnarray}  
This yields, for example, 
\begin{eqnarray}
 \int_{-\infty}^{\infty} d\nu e^{i\nu\eta}h_{i\nu}(k_{\bot}\tau) =
\int_{-\infty}^{\infty} {d~\theta \over \cosh^2\theta}
\sin [k_{\bot}\tau\cosh(\theta-\eta)]~~.               
\label{eq:A3.7}\end{eqnarray}  
After integration over $\nu$ we obtain the following integral for
$\Delta^{(L)}_{rs}$ ,
\begin{eqnarray}
 \Delta^{(L)}_{rs}= \int {d^2{\vec k}\over (2\pi)^3}
{k_rk_s\over  k_{\bot}^2} e^{i{\vec k}{\vec r}}     
\int_{0}^{\tau_2}{d\tau\over\tau}
\bigg( \pi k_{\bot}\tau\cosh\eta e^{-k_{\bot}\tau\sinh |\eta|}
- \int_{-\infty}^{\infty} {d~\theta \over \cosh^2\theta}
\sin [k_{\bot}\tau\cosh(\theta-\eta)]\bigg)~,               
\label{eq:A3.8}\end{eqnarray}  
and similar integrals for the other components. 
The first term in this formula is calculated in the following way.
After integration over $\tau$ we continue: 
\begin{eqnarray} 
 \Delta'_{rs}= - {\partial_r\partial_s \over 8 \pi^2} \coth |\eta|~ 
 \int {d^2{\vec k}\over  k_{\bot}^2} e^{i{\vec k}{\vec r}}     
[1- e^{-k_{\bot}\tau_2\sinh |\eta|}]  \hspace{6cm} \nonumber\\
= - {\partial_r\partial_s \over 4 \pi}  \coth |\eta|   
\int_{0}^{\infty} {d k_{\bot}\over k_{\bot}}  J_0(k_{\bot}r_{\bot})
[1-e^{-k_{\bot}\tau_2\sinh |\eta|}]=
- {\partial_r\partial_s \over 4 \pi}  \coth |\eta|  
\ln\bigg[ { \tau_2\sinh |\eta| + \sqrt{ r_{\bot}^2+\tau_{2}^{2}\sinh^2\eta }
\over r_{\bot}}\bigg]~~.                           
\label{eq:A3.9}\end{eqnarray}                             
To work out the second term, one should introduce $k_z=k_{\bot}\sinh\theta$
and  $k_0=k_{\bot}\cosh\theta=|{\bf k}|$ and change $d^2{\vec k}d\theta$ for
the three-dimensional integration $d^3{\bf k}$. 
With $t=\tau\cosh\eta$, ${\bf r}=(x,y,\tau\sinh\eta)$, this leads to
\begin{eqnarray} 
 \Delta''_{rs}=  {\partial_r\partial_s \over {2 \pi}^3} 
\int_{0}^{\tau_2}{d\tau\over\tau}
 \int {d^3{\bf k}\over  k_{0}^{3}} e^{i{\bf k}{\bf r}}\sin k_0 t
={\partial_r\partial_s \over 4\pi} \int_{0}^{\tau_2}{d\tau\over\tau}
\bigg( \theta(r_{\bot}^2-\tau^2)
{\tau\cosh\eta\over\sqrt{r_{\bot}^2+\tau^{2}\sinh^2\eta }} 
+ \theta(\tau^2-r_{\bot}^2) \bigg) \nonumber\\
= {\partial_r\partial_s \over 4\pi}
\bigg(\theta(r_{\bot}-\tau_2)~ \coth |\eta|~  
\ln\bigg[ { \tau_2\sinh |\eta| + \sqrt{ r_{\bot}^2+\tau_{2}^{2}\sinh^2\eta }
\over r_{\bot}}\bigg]  +\theta(\tau_2-r_{\bot})~
\ln {\tau_2\over r_{\bot}}\bigg)~~.        
\label{eq:A3.10}\end{eqnarray}                             
Adding (\ref{eq:A3.9}) and (\ref{eq:A3.10}), we obtain the first of the 
equations (\ref{eq:E5.1}). 

\bigskip                  

\renewcommand{\theequation}{D\arabic{equation}}
\setcounter{equation}{0}  

\centerline {\bf APPENDIX ~D. Gluon correlators in the central rapidity 
          region and  near the light wedge} 
\bigskip

In this section, we compare the correlators  of the gauge $A^\tau=0$ with
the similar correlators in the  three other gauges, $A^0=0$, $A^+=0$ and
$A^-=0$.  We shall start with the simplest on-mass-shell Wightman 
function $\Delta_{10}^{\mu\nu}$. These type of correlators, 
$\Delta_{01}^{\mu\nu}$, $\Delta_{0}^{\mu\nu}$ and $\Delta_{1}^{\mu\nu}$
share the same polarization sum of the free gauge field. They correspond
to the densities of the final states of the  radiation field and are
important for various calculations.  The same polarization  sum appears in
expressions for the transverse part of the propagators,
$\Delta_{ret}^{\mu\nu}$, $\Delta_{adv}^{\mu\nu}$, $\Delta_{00}^{\mu\nu}$ 
and $\Delta_{11}^{\mu\nu}$. For our immediate purpose we shall include the
projector $d^{\mu\nu}$ of the gauge $A^\tau=0$ to the formal Fourier
representation,
\begin{eqnarray}
i D^{\mu\nu}_{10}(x_1,x_2)= 
\int {d^3 k\over (2\pi)^3 2k^0}  d^{\mu\nu}(k; x_1,x_2) 
e^{-ik(x_1-x_2)}, 
\label{eq:E4.1} 
\end{eqnarray}        
with the ``extraneous'' dependence of the Fourier transform on the
time and spatial coordinates.  This dependence disappear 
in some important limits.  Therefore, we discover the domains
where the wedge dynamic simplifies and describes the
processes which are approximately homogeneous in space and time.
These domains are: (i) the central rapidity region, $\eta_{1,2}\ll 1$ (or
$x^{3}_{1,2}\sim 0$), where the projector in the integrand of
Eq.~(\ref{eq:E4.1}) is
\begin{eqnarray}
 d^{\mu\nu}(k,u)= -{\rm g}^{\mu\nu} +
{k^\mu u^{\nu}+u^\mu k^{\nu} \over  k u } - {k^\mu k^{\nu}\over (k u)^2 }~, 
\label{eq:E4.2} 
\end{eqnarray}       
with the gauge-fixing vector, $u^\mu =(1,0,0,0)$, which approximately 
coincide with the local normal to the hypersurface $\tau =const$;
and (ii), the vicinities of two null-planes, $\eta\rightarrow\pm\infty$
(or $x^\mp\rightarrow 0$) where 
\begin{eqnarray}
 d^{\mu\nu}(n_{\pm},k)= -{\rm g}^{\mu\nu} +
{k^\mu n^{\nu}_{\pm}+k^\nu n^{\mu}_{\pm} \over  (k n_{\pm}) }
\label{eq:E4.3} ~,
\end{eqnarray}       
with the null-plane vectors ~$n^{\mu}_{\pm}=(1,0,0,\mp 1)~$. 

Eqs.~(\ref{eq:E3.5})-(\ref{eq:E3.8}) almost fit our needs.
In all three cases ($x^3\rightarrow 0$, $k^0x^0\gg 1$,  as well as 
$x^-\rightarrow 0$,  $k^-x^+ \gg 1$  and 
$x^+\rightarrow 0$,  $k^+x^- \gg 1$)  the functions $f_1$ and $f_2$
can be approximated by the following expressions:
\begin{eqnarray}
f_{1} \approx 
 i~ \tanh (\theta-\eta) e^{-ik_{\bot}\tau \cosh (\theta-\eta)
+i{\vec k}{\vec r} } =
{k^0x^3-k^3x^0 \over k^0x^0-k^3x^3}e^{-ikx}=
{k^+x^- -k^-x^+ \over k^+ x^- +k^- x^+}e^{-ikx}~, 
\label{eq:E4.4}
\end{eqnarray}         
\begin{eqnarray}
f_{2} \approx 
 ik_{\bot}\tau~ {e^{-ik_{\bot}\tau \cosh (\theta-\eta)
+i{\vec k}{\vec r} } \over \cosh (\theta-\eta)}=
ik_{\bot}^{2}\tau^2 {e^{-ikx}\over k^0x^0-k^3x^3}=
2ik_{\bot}^{2}\tau^2 {e^{-ikx} \over k^+ x^- +k^- x^+}~~. 
\label{eq:E4.4a}
\end{eqnarray}          
(We have omitted the time independent terms in $f_1$ and $f_2$ which
set the potentials of the mode $v^{(2)}$ to zero at $\tau=0$. These
kind of terms would correspond to the residual gauge symmetry and
is not kept in the axial and the null-plane gauges as well. Thus, we cannot
really claim the correspondence of the longitudinal fields between the wedge
dynamics and these three dynamics.)

Transformation of the correlator $\Delta^{lm}(1,2)$ to the Minkowski
coordinates is carried out according to the formula,
\begin{equation}
D^{\mu\nu}(x_1,x_2)=a^{\mu}_{i}(x_1){\rm g}^{il}(x_1) 
\Delta_{lm}(u_1,u_2){\rm g}^{mk}(x_2) a^{\nu}_{m}(x_2)~,  
\label{eq:E4.5}\end{equation}  
where the matrix of the transformation is defined in the standard way,
\begin{equation}
a^{\mu}_{i}(x) = {\partial x^\mu \over \partial u^i},~~~
a^{0}_{\eta}(x) = x^3,~~~a^{3}_{\eta}(x) = x^0,
~~~a^{r}_{s}=\delta^{r}_{s}~~.
\label{eq:E4.6}\end{equation} 
These are the only components of the tensor $a^{\mu}_{i}(x)$ which
participate in the transformation.  In this way, we obtain,
\begin{eqnarray}
D^{00}(1,2)=x^{3}_{1}x^{3}_{2}\Delta^{\eta\eta}(1,2);\;\; 
D^{03}(1,2)=x^{3}_{1}x^{0}_{2}\Delta^{\eta\eta}(1,2) \nonumber \\
D^{30}(1,2)=x^{0}_{1}x^{3}_{2}\Delta^{\eta\eta}(1,2);\;\; 
D^{33}(1,2)=x^{0}_{1}x^{0}_{2}\Delta^{\eta\eta}(1,2), \nonumber \\
D^{0r}(1,2)=x^{3}_{1} \Delta^{\eta r}(1,2);\;\; 
D^{r0}(1,2)=x^{3}_{2} \Delta^{r\eta }(1,2);\nonumber \\ 
D^{3r}(1,2)=x^{0}_{1} \Delta^{\eta r}(1,2);\;\; 
D^{r3}(1,2)=x^{0}_{2}\Delta^{r\eta}(1,2),\nonumber\\
D^{rs}(1,2)=\Delta^{rs}(1,2)~~. 
\label{eq:E4.7}\end{eqnarray}                              
Every additional factor ${\rm g}^{\eta\eta}=\tau^{-2}$ finds a counterpart
which prevents singular behavior at $\tau =0$. In the above
approximation, the expression for the  $\Delta^{\eta\eta}(x_1,x_2)$
component of the correlator has the form,
\begin{eqnarray}
\Delta^{\eta\eta}(x_1,x_2)= 
\int {d^3 k\over (2\pi)^3 2k^0} {k_{\bot}^{2}e^{-ik(x_1-x_2)}
\over (k^0x_{1}^{0}-k^3x_{1}^{3})(k^0x_{2}^{0}-k^3 x_{2}^{3})}  
=\int {d^3 k\over (2\pi)^3 2k^0} {4 k_{\bot}^{2}e^{-ik(x_1-x_2)}
\over (k^+x_{1}^{-}+k^-x_{1}^{+})(k^+x_{2}^{-}+k^- x_{2}^{+})}~.  
\label{eq:E4.8} 
\end{eqnarray}  
Therefore, in the limit of $x^{3}_{1,2}\rightarrow 0$ we obtain that
$D^{00},D^{0i}\rightarrow 0$, while $d^{33}(k,u)\rightarrow 
k_{\bot}^{2}/k_{0}^{2}$, thus reproducing the corresponding components
of the gauge $A^0=0$. The other components are reproduced one by one as well, 
and one can expect a smooth transition between the gauge of the the wedge 
dynamic and the local temporal axial gauge of the  reference frame  co-moving
with the dense quark-gluon matter created in the collision.
 
In the limits of $x^{\mp}_{1,2}\rightarrow 0$ we obtain that
\begin{eqnarray}
D^{00},D^{03},D^{30},D^{33}\rightarrow {k_{\bot}^{2}\over (k^{-})^{2}}
={k^+ \over k^-}~,~~~ {\rm if}~~x^-\rightarrow 0~~,    
\label{eq:E4.9} 
\end{eqnarray}  
and 
\begin{eqnarray}
D^{00},D^{03},D^{30},D^{33}\rightarrow {k_{\bot}^{2}\over (k^{+})^{2}}
={k^- \over k^+}~,~~~ {\rm if}~~x^+\rightarrow 0~~.    
\label{eq:E4.10} 
\end{eqnarray}  
These limits, after they are found for all components, lead to the well
known  expressions of the projectors $d^{\mu\nu}(n,k)$ in the null-plane
gauge (the gauge $A^{+}=0$ in the vicinity of $x^+=0$, and  $A^{-}=0$ in
the vicinity of $x^-=0$). Therefore we obtained an expected result; in the
limit of the light-cone  propagation, the gauge $A^{\tau}=0$ recovers the
null-plane gauges $A^{+}=0$ and $A^{-}=0$. 
 
Some remarks are in order. First, the concept of the structure
functions relies heavily  on the null-plane dynamics which essentially
uses these gauges. For two hadrons (or two nuclei) we have two different
null-plane dynamics which do not share the same Hilbert space of states.
Now we have an important opportunity to describe both nuclei and the 
fields produced in their interaction within the same dynamic and the same 
Hilbert space. Second, one may trace back the origin of the poles $(ku)^{-1}$
in the polarization sums of axial gauges  $(uA)=0$ and see
that they appear in the course of the approximation of the less-singular
factor  $[k_\bot\cosh(\theta-\eta)]^{-1}$ in various limits of the propagator
of the gauge $A^\tau=0$.

Further, contrary to the naive expectation 
that we obtain the gauge $A^{+}=0$ at
$x^{-}=0$ and the gauge $A^{-}=0$ at $x^{+}=0$, we obtained them in the
opposite correspondence. First of all, let us notice that the result is
mathematically consistent. Indeed, the gauge condition  $A^{\tau}=0$ 
may be rewritten in the form, 
\begin{eqnarray} 
A^{\tau}={1\over 2}(A^+ e^{-\eta} + A^- e^{\eta} )=0~~. 
\label{eq:E4.11}  \end{eqnarray}     
Thus the limit of $\eta\rightarrow\infty$ ($x^{-}\rightarrow 0$) indeed 
leads to $A^{-}=0$ and the limit of  $\eta\rightarrow - \infty$
($x^{+}\rightarrow 0$) leads to $A^{+}=0$ as the limiting gauge conditions.
Recalling that
\begin{eqnarray} 
A^{\eta}={1\over 2}(A^+ e^{-\eta} - A^- e^{\eta} )= A^+
e^{-\eta} = -A^- e^{\eta} ~~, 
\label{eq:E4.12}  \end{eqnarray}  
we immediately realize that in the vicinities of both null-planes, the 
tangent component $A^{\eta}=0$. This fact has a very simple geometrical
explanation; the normal and tangent vectors of the null plane are
degenerate. Once  $A^{\tau}=0$, we have $A^{\eta}=0$ and consequently,
$A^{+}=0$ and $A^{-}=0$ at  $\eta\rightarrow\pm\infty$.  This result
naturally follows from the geometry of the system of the surfaces where we
define the field states. These are subject to dynamical evolution in the
direction  which is normal to the hyper-surface.

\end{document}